\documentclass[preprint,pre,eqsecnum,longbibliography]{revtex4-1} 

\usepackage{natbib}
\usepackage{graphicx}
\usepackage{subfigure}
\usepackage{amssymb}
\usepackage{amsfonts}
\usepackage{amsmath}
\usepackage{amsbsy}
\usepackage{mathrsfs} 
\usepackage{color}
\usepackage{gensymb} 
\usepackage{hyperref} 
\usepackage{soul}
\usepackage{lineno}
\usepackage{esint}
\usepackage{float}
\usepackage{mathtools}
\providecommand\bnabla{\boldsymbol{\nabla}}

\providecommand\bcdot{\boldsymbol{\cdot}}

\providecommand\bx{\mathbf{x}}

\providecommand\bE{\mathbf{E}}

\providecommand\be{\mathbf{\hat{e}}}
\providecommand\bn{\mathbf{\hat{n}}}

\providecommand\unit{\boldsymbol{\hat{\imath}}}

\newcommand{\pd}[2]{\frac{\partial #1}{\partial #2}}

\newcommand{\ub}[1]{^{({#1})}}

\begin{document}
\title{Plasmonic resonances of slender nanometallic rings}
\author{Matias Ruiz}
\affiliation{School of Mathematics, University of Edinburgh, Edinburgh EH9 3FD, UK}
\author{Ory Schnitzer}%
\affiliation{%
Department of Mathematics, Imperial College
London, London SW7 2AZ, UK
}%
\begin{abstract}
We develop an approximate quasi-static theory describing the low-frequency plasmonic resonances of slender nanometallic rings and configurations thereof. First, we use asymptotic arguments to reduce the plasmonic eigenvalue problem governing the geometric (material- and frequency-independent) modes of a given ring structure to a 1D-periodic integro-differential problem in which the eigenfunctions are represented by azimuthal voltage and polarization-charge profiles associated with each ring. Second, we obtain closed-form solutions to the reduced eigenvalue problem for azimuthally invariant rings (including torus-shaped rings but also allowing for non-circular cross-sectional shapes), as well as coaxial dimers and chains of such rings. For more general geometries, involving azimuthally non-uniform rings and non-coaxial structures, we solve the reduced eigenvalue problem using a semi-analytical scheme based on Fourier expansions of the reduced eigenfunctions. Third, we used the asymptotically approximated modes, in conjunction with the quasi-static spectral theory of plasmonic resonance, to study and interpret the frequency response of a wide range of nanometallic slender-ring structures under plane-wave illumination.
\end{abstract}

\maketitle

\section{Introduction}
\label{sec:intro}

Localized-surface-plasmon resonance, namely the excitation of collective electric-field and electron-charge-density oscillations in metallic nanoparticles and nanostructures, has been instrumental over the past several decades in enabling new techniques for manipulating visible and near-infrared electromagnetic waves on nanometric scales, with applications including sensing, targeted heating and metamaterials \cite{Maier:07, Klimov:14}. An important motif in fundamental research as well as applications has been the use of nearly singular, i.e., multiple-scale, geometries to achieve tunable and high quality resonances, field enhancement in nanoscale `hotspots' and for probing light-matter interactions and non-classical phenomena \cite{Romero:06, Schuller:10, Suh:2012, Chen:13, Toscano:13}. 

The class of multiple-scale nanometallic structures most extensively studied in nanoplasmonics consists of closely spaced particles and similarly particles very near to a metallic substrate \citep{Tamaru:02,Gunnarsson:05, Jain:07, Klimov:07,Aubry:10A,Aubry:11,Lebedev:13,Pendry:13,Kadkhodazadeh:14,Schnitzer:15plas,Schnitzer:16,Schnitzer:16b,Schnitzer:18,Yu:18,Yu:19}. Such structures exhibit a rich spectrum comprised of several distinct families of modes characterized by their symmetries, as well as behavior in the limit where the aspect ratio $\kappa$, say the ratio of particle radius to gap width, is large. Particular emphasis has been given to bonding-gap modes, which are characterized by localization of the electric field to the gap, with the field directed across the gap. For large $\kappa$, the characteristic angular frequencies $\omega$ associated with these modes behave roughly like $\omega/\omega_p\simeq1/\kappa^{1/4}$, wherein $\omega_p$ denotes the plasma angular frequency \cite{Schnitzer:15plas}. For large $\kappa$, the first few bonding-gap modes strongly couple with incident radiation, giving rise to low-frequency resonances associated with a giant field enhancement in the gap. 

Another class of multiple-scale nanometallic structures that is commonly employed consists of slender particles such as high-aspect-ratio nanorods and spheroids \citep{Link:99,Sonnichsen:02,Prescott:06,Khlebtsov:07,Neubrech:08,Lu:11,Agha:14, Deng:2020}. Parallels can be drawn between the bonding-gap modes of closely spaced particles and the longitudinal modes of slender particles, which are characterized by polarization-charge distributions that for sufficient slenderness vary mainly along the particle axis. The latter longitudinal modes redshift as the particle becomes more and more slender, analogously to the redshift of the bonding-gap modes described above. In the slender-particle case, however, the plasmon-frequency redshift is far more singular: $\omega/\omega_p\simeq1/\kappa$, where now $\kappa$ represents the ratio of length to thickness \cite{Prescott:06}. Accordingly, the first few longitudinal modes of a slender particle typically give rise to remarkably low-frequency plasmonic resonances (down to the near-infrared regime), which are typically associated with high quality factors and distinctive directional characteristics. 

In comparison with straight slender particles, curved slender particles such as slender rings and helixes are more subwavelength at their longitudinal resonances. Indeed, let the length and thickness of some slender particle be $2l$ and $2b$, respectively; the scaling $\omega/\omega_p\simeq 1/\kappa$ implies longitudinal resonances at wavelengths $\lambda$ roughly proportional to $(l/b) \lambda_p$, where $\lambda_p=2\pi c/\omega_p$ and $c$ is the speed of light in vacuum. Hence, if the characteristic linear dimension of the particle is $2a$, we have that $a/\lambda\simeq (a/l)(b/\lambda_p)$. One avenue to reducing the latter size factor is to reduce $b$, though this is ultimately limited by manufacturing capabilities as well as non-classical effects at subnanometric scales. An alternative avenue is to coil the centerline so as to decrease $a/l$. In particular, $a/l$ is unity for a straight particle and $1/\pi$ for a circular ring.

The above characteristics of high-aspect-ratio nanometallic particles can be indirectly inferred from classical analytical solutions for cylinder and sphere dimers, ellipsoids and tori, which are exact in the quasi-static approximation \cite{Moussiaux:1977, Aizpurua:03, Mary:05, Dutta:08,Bohren:Book, Klimov:14, Voicu:17, Downing:2020}. While these solutions allow for arbitrary aspect ratio, they are limited to separable geometries. Even then, exact solutions tend to be cumbersome and sometimes it is not straightforward to extract singular behaviors in the high-aspect-ratio limit. For more general geometries, analytical solutions are unavailable, whereas brute-force finite-element simulations become expensive at high aspect ratios. An alternative, more insightful and versatile, theoretical approach is to consider the high-aspect-ratio limit --- in which plasmonic resonance is expected to be most pronounce --- from the outset, using tools of asymptotic analysis and singular-perturbation theory \cite{Hinch:91}. The latter tools are classical, though their application to plasmonics is relatively new and so far the focus has been on near-touching geometries \cite{Schnitzer:15plas, Schnitzer:18}. 

Recently, we took a first step in applying an asymptotic approach to the analysis of the plasmonic properties of slender nanometallic structures \cite{Ruiz-Schnitzer:19}. In particular, we developed an approximate theory of the longitudinal plasmonic resonances of slender bodies of revolution, with the thickness profile along the symmetry axis being essentially arbitrary. Besides key scalings, new closed-form approximations for special geometries and physical insight into the role of slenderness and shape, this work furnished a versatile semi-analytical scheme which allows one to rapidly calculate the plasmonic response of a certain family of complex 3D geometries. 

The framework in \cite{Ruiz-Schnitzer:19} is underpinned by a spectral theory which is exact in the quasi-static approximation; it is based on the so-called plasmonic eigenvalue problem \cite{Mayergoyz:05, Grieser:14, Klimov:14, Davis:17}, which defines a set of material- and frequency-independent modes of a nanometallic structure based on its geometry, with the relative permittivity of the structure playing the eigenvalue role. Given this set of modes, calculating the optical response of the structure for arbitrarily prescribed radiation sources and physical values of the particle permittivity amounts to the evaluation of normalization and overlap integrals involving the modes and the incident radiation field. At near-resonance frequencies, often a single mode dominates the spectral expansion; this modal  approach  is  accordingly both efficient and insightful in that it provides a linkage between physical resonances and mathematical eigenfunctions.

In \cite{Ruiz-Schnitzer:19}, we used asymptotic tools to calculate modes as well as overlap integrals. In particular, to analyze the plasmonic eigenvalue problem, we relied mainly on the method of matched asymptotic expansions, where the physical domain is decomposed into distinguished regions which are separately analyzed and then matched together; the use of such methodology in the context of slender particles is especially popular in fluid dynamics, where it is known as `slender-body theory' \citep{Tillett:70,Batchelor:70,Cox:70, Keller:76, Van:pert,Handelsman:67}. This allowed us to reduce the plasmonic eigenvalue problem for the body of revolution (restricted to the longitudinal --- in this case axisymmetric --- modes) to an asymptotically equivalent 1D problem whose domain is a finite line segment corresponding to the body's centerline. In this reduced problem, the eigenvalue is the original permittivity eigenvalue scaled by $\kappa^2$. With that rescaling, the reduced problem involves $\kappa$ only weakly, namely through its logarithm. The eigenfunctions in the reduced problem represent effective 1D voltage and polarization-charge profiles. From these profiles, the corresponding mode distributions can be evaluated in 3D, everywhere in the interior and exterior of the particle; they can also be used to easily evaluate the normalization and overlap integrals appearing in the spectral formulation. The reduced eigenvalue problem consists of a differential equation which is coupled to an integral equation.  The differential equation represents an effective Gauss law for an infinitesimal longitudinal segment of the body. (The boundary conditions are dependent on the local geometry of the tips.) The integral equation represents a spatially nonlocal capacity relation accounting for electrostatic interactions between different longitudinal segments of the body. 

The purpose of the present work is to extend the plasmonic slender-body in \cite{Ruiz-Schnitzer:19} to slender nanometallic rings and configurations thereof. By a slender ring we mean a body whose thickness about a circular curve is small relative to the radius of that circle. We shall allow azimuthal variations in thickness and non-circular cross-sectional shapes, thus going considerably beyond the idealized case of a slender torus; we shall also consider ring dimers and chains, the separation between the rings being comparable to their radii. An obvious source of motivation for focusing on ring structures is that they are commonly employed in nanoplasmonics \cite{Nordlander:09}; they are moderately compact, as discussed above, relatively easy to fabricate \cite{Chow:20}, their hollow center is advantageous for sensing applications \cite{Aizpurua:03,Aziz:2017} and offer increased tuning through aspect ratio \cite{Aizpurua:03,Nordlander:09}, azimuthal thickness profile \cite{Bochenkov:17,Zong:19} and  multiple-ring interactions \cite{Zhang:15,Pattanayak:17,Zhang:19,Dana:20}. We also have technical reasons to focus on ring geometries. Thus, the absence of tips  eliminates the need to derive boundary conditions to close the reduced eigenvalue problem, which is generally quite a subtle aspect of the theory. Furthermore, the fact that rings degenerate to circular curves in the high-aspect-ratio limit will be seen to enable convenient diagonalizations of the integral operators appearing in the reduced problem. 

The rest of the paper is structured as follows. In \S\ref{sec:QS}, we formulate the general problem of scattering of a plane wave from a nanometallic structure in the quasi-static approximation and then review the plasmonic eigenvalue problem and its use for solving that scattering problem. In \S\ref{sec:long}, we employ asymptotic arguments in the high-aspect-ratio limit to derive a reduced eigenvalue problem in the case of a single ring; at this stage, we assume circular cross sections, although the thickness may vary in the azimuthal direction. To emphasize the physics, the asymptotic arguments are described intuitively and in dimensional notation; readers requiring a more formal justification of the approximations can refer to the derivation of the reduced problem in \cite{Ruiz-Schnitzer:19} and textbooks describing the method of matched asymptotic expansions and slender-body theory \cite{Hinch:91}. In the remainder of \S\ref{sec:long}, we discuss the singular scaling of the permittivity eigenvalues, the accuracy of the approximation scheme and develop closed-form solutions for torus-shaped rings and a semi-analytical scheme for azimuthally non-uniform rings. In \S\ref{sec:dimer}, we generalize the reduced eigenvalue problem to the case of ring dimers, develop closed-form solutions for coaxial dimers formed of torus-shaped rings and a semi-analytical scheme for more general dimer geometries. In \S\ref{sec:extensions}, we briefly discuss further geometrical extensions to non-circular cross sections and chains of rings; the latter extension is illustrated by considering the consequences of a defect in a  coaxial chain of ring dimers. In \S\ref{sec:PW}, we demonstrate the application of the asymptotic modes to calculate absorption cross sections for ring configurations illuminated by a plane wave. In \S\ref{sec:qsvalid}, we discuss limits of validity of the quasi-static approximation specifically in describing the longitudinal resonances of slender-ring structures, and show comparisons with full-wave simulations. We give concluding remarks in \S\ref{sec:concluding}.

\section{Quasi-static formulation}\label{sec:QS}
\subsection{Scattering problem}\label{ssec:qs}
Consider a homogeneous nanometallic structure in vacuum. We assume that the characteristic linear dimension of the structure is sufficiently small relative to the free-space wavelength such that a quasi-static approximation is applicable (\cite{Bohren:Book}, cf.~\S\ref{sec:qsvalid}). In what follows, we formulate a scattering problem governing the electric near-field in the vicinity of the structure in the scenario where the structure is illuminated by a plane wave; for simplicity, we do not include in the formulation the possibility of near-field external sources. 

In the quasi-static approximation, the electric field is irrotational in the vicinity of the structure. Defining this near-field to be the real part of $\bE e^{-i\omega t}$, wherein $\omega$ is angular frequency and $t$ time, we may accordingly introduce an electric potential $\varphi$ such that $\bE=-\bnabla\varphi$. Given the absence of near-field external sources, the potential $\varphi$ satisfies 
\begin{equation}\label{QS problem}
	\bnabla \bcdot \left(\epsilon\bnabla\varphi\right)=0,
\end{equation} 
where $\epsilon$ denotes the permittivity relative to vacuum. In the exterior of the structure, $\epsilon=1$. In the interior of the  structure, $\epsilon$ is given by the  complex-valued and frequency dependent relative permittivity of the metal, say $\epsilon_r(\omega)$, for which empirical data is available (e.g., \citep{Johnson:72}). Alternatively, we may refer to the Drude model \cite{Maier:07}
\begin{equation}\label{drude}
	\epsilon_r(\omega) = 1- \frac{\omega_p^2}{\omega^2+i\gamma \omega},
\end{equation}
wherein $\omega_p$ is the plasma frequency and $\gamma$ is a parameter representing ohmic losses. The Drude model is adequate for frequencies considerably lower than $\omega_p$, which as we shall see is the case for the longitudinal resonances of slender-ring structures. This model will be used for the sake of illustration in  \S\ref{sec:PW}. 

The quasi-static formulation is closed by the far-field condition 
\begin{equation}\label{QSfar}
	\bE\to\bE_{\infty} \quad \text{as} \quad |\bx|\to\infty,
\end{equation}
where the constant vector $\bE_{\infty}$ represents the incident plane-wave. Specifically, keeping in mind that the near-field domain is small relative to the free-space wavelength, $\bE_{\infty}$ is the electric-field phasor of the incoming wave evaluated at the location of the structure.

\subsection{Plasmonic eigenvalue problem}\label{ssec:PEP}
Upon setting $\bE_{\infty}$ to zero, \eqref{QS problem} and \eqref{QSfar} give the so-called  `plasmonic eigenvalue problem' governing the localized-surface-plasmon resonances of the nanometallic structure. Since frequency enters this problem solely via the metal permittivity $\epsilon_r$, it is natural to take $\epsilon_r$ as the eigenvalue, henceforth denoted $\mathcal{E}$, rather than the frequency. The permittivity eigenvalues are scale invariant and independent of material and frequency; they are determined solely by the geometry of the structure. For a smooth geometry, there are infinitely many negative-real permittivity eigenvalues which accumulate at $-1$ \cite{Mayergoyz:05, Klimov:14,Grieser:14, Ando:16}. The associated field distributions, namely eigenfunctions, are also scale-invariant up to a linear stretching and can always be chosen real-valued. 

Physically, the eigenfunctions can be interpreted as perpetual localized-surface-plasmon oscillations in the absence of material losses and external forcing. (Radiation losses are effectively neglected in the quasi-static approximation.) Mathematically, the set of eigenfunctions possesses completeness and orthogonality properties that can be used to explicitly solve arbitrary scattering problems. In particular, the solution to the scattering problem formulated above can be written as the spectral expansion
(see \citep{Bergman:79,Klimov:14,Grieser:14,Ammari_book:18,Davis:17})
\begin{equation}
	\label{Field expansion}
	\varphi(\mathbf{x}) = -\mathbf{E}_{\infty}\boldsymbol{\cdot}\mathbf{x} + \sum_{I \in \mathcal{I}}\frac{\epsilon_r(\omega)-1}{\epsilon_r(\omega)-\mathcal{E}^{(I)}}\frac{\int dV\, 
		\mathbf{E}_{\infty}
		\boldsymbol{\cdot} \boldsymbol{\nabla}\varphi\ub{I}}{\int dV\, \boldsymbol{\nabla}\varphi\ub{I} \boldsymbol{\cdot} \boldsymbol{\nabla}\varphi\ub{I}}\,\varphi\ub{I}(\mathbf{x}),
\end{equation} 
where $\mathcal{I}$ is an index set for the eigenvalues and eigenfunctions, $dV$ is a volume element and the integrals are over the interior of the structure. The integrals in the numerator and denominator are called overlap and normalization integrals, respectively. We note that scattering problems involving near-field sources can be treated similarly, with $\mathbf{E}_{\infty}$ replaced by the total field in the absence of the structure. From \eqref{Field expansion}, quasi-static approximations for the optical cross sections can be readily calculated, as we shall see in \S\ref{sec:PW}. 

For frequencies such that the complex-valued permittivity $\epsilon_r(\omega)$ is close to a permittivity eigenvalue, the corresponding term in the expansion \eqref{Field expansion} --- or terms, in the case of perfect or near-perfect degeneracy --- may become dominant. In that case, we say that the associated eigenfunctions (or eigenmodes) are resonantly excited by the incident radiation. Thus, for such near-resonance frequencies, the spectral solution can be asymptotically simplified, revealing a connection between the physical and mathematical perspectives described above. The existence and characteristics of such resonances in practice, however, depend on several additional factors including the relative smallness of the imaginary component of $\epsilon_r(\omega)$, the overlap between the incident field and the eigenfunctions participating in the resonance, as well as interference effects. All of these dependencies can be studied using \eqref{Field expansion} once the eigenvalues and eigenfunctions of the geometry have been calculated.

\section{Longitudinal modes of a single slender ring}\label{sec:long}
\subsection{Geometry}\label{ssec:geometry}
\begin{figure}[t!]
	\begin{center}
		\includegraphics[scale=0.5]{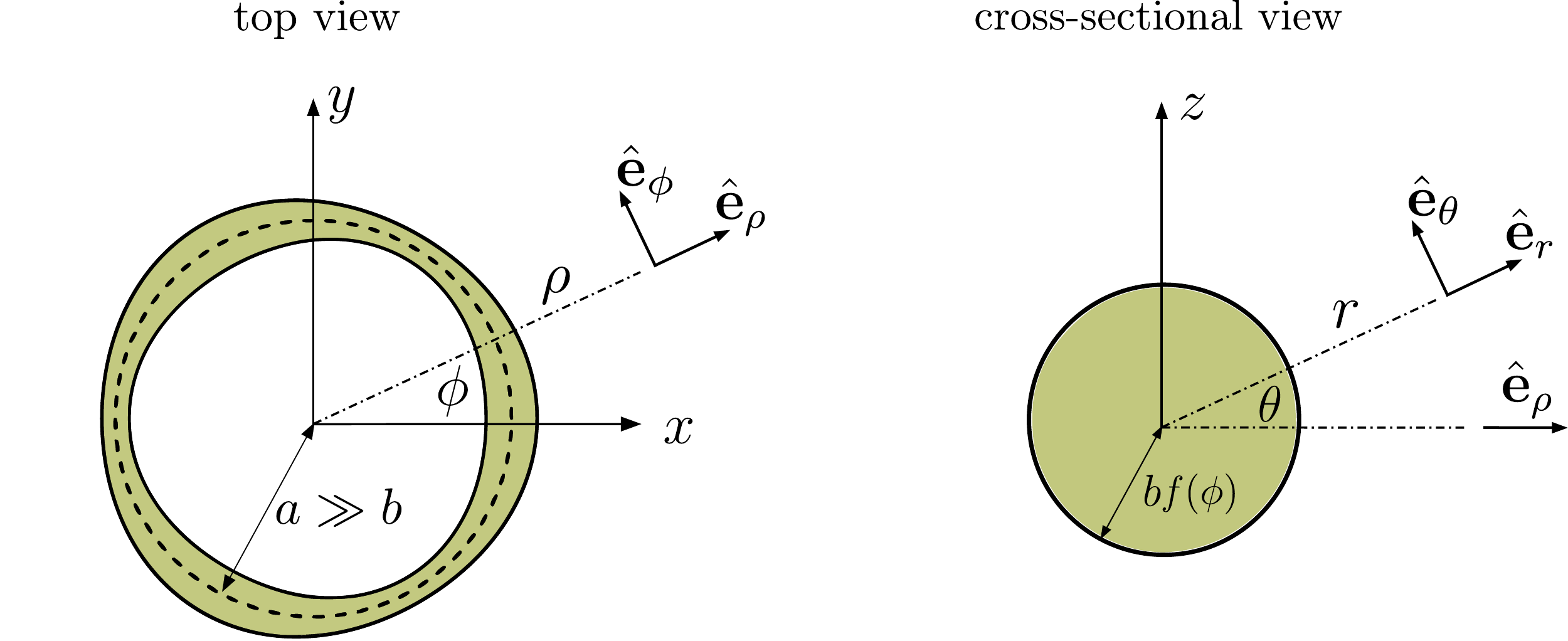}
		\caption{Schematic of a slender ring with circular cross sections as considered in \S\ref{sec:long}.}
		\label{fig:schematic1}
	\end{center}
\end{figure}

In this section we consider the plasmonic eigenvalue problem (cf.~\S\ref{ssec:PEP}) in the case of a single nano-sized ring of radius $a$ and characteristic thickness $b$. For now, we assume that the ring is formed of circular cross sections of radius $bf(\phi)$, which are centered about a circular centerline of radius $a$; the thickness profile $f(\phi)$ is a dimensionless function of the azimuthal angle $\phi$. (Non-circular cross-sections are treated in \S\ref{ssec:noncirc}.) In Fig.~\ref{fig:schematic1}, we define  the Cartesian $(x,y,z)$, cylindrical $(\rho,z,\phi)$ and `cross-sectional' $(r,\theta,\phi)$ coordinate systems. Accordingly, the position measured from the center of the ring can be written $\bx=\mathbf{y}+r\be_r$, where $\be_r=\be_\rho\cos\theta+\be_z\sin\theta$ and $\mathbf{y}(\phi)=a\be_{\rho}$ is the circular centerline, with $\be_{\rho}=\be_x\cos\phi+\be_y\sin\phi$. Furthermore, the ring boundary can be written $r=bf(\phi)$. 

\subsection{Reduced eigenvalue problem} \label{sec:reduced evp}
Henceforth, we consider slender rings for which the aspect ratio $\kappa=a/b$ is large and $f(\phi)$ is of order unity. In particular, our interest is in modes, i.e., solutions of the plasmonic eigenvalue problem, for which $\mathcal{E}\to-\infty$ as $\kappa\to\infty$. (According to the Drude model \eqref{drude}, large-negative permittivity eigenvalues imply low frequencies $\omega\ll\omega_p$.) For any mode with this asymptotic property, scaling arguments suggest that, as $\kappa\to\infty$, cross-sectional variations of the interior potential become negligible relative to azimuthal variations. We accordingly label such modes  `longitudinal'. We note that in the case of a body of revolution \cite{Ruiz-Schnitzer:19}, those modes can be identified based on symmetry as they simply correspond to the axisymmetric modes.

In light of the above, we approximate the interior potential as
\begin{equation}\label{v def}
	\varphi(\bx)= v(\phi),
\end{equation}
where the azimuthal function $v(\phi)$ represents the voltage relative to infinity. (Without loss of generality, we take the potential to vanish at infinity.)
Another key azimuthal function is the polarization-charge line density 
\begin{equation}\label{q def}
	q(\phi)=\oint\limits_{r=bf(\phi)} dl\,\sigma,
\end{equation}
where $\sigma=\epsilon_0[\bE\bcdot\bn]_i^e$ is the polarization-charge surface density, $dl$ is a differential length element and the integral is over the circular cross-sectional interface $r=bf(\phi)$ at constant $\phi$; in the expression for $\sigma$, $\epsilon_0$ is the permittivity of vacuum, $\bn$ an outward normal unit vector and the square brackets stand for the jump across the interface in that direction, the subscript $i$ and superscript $e$ indicating the interior and exterior sides, respectively. For $|\mathcal{E}|\gg1$, continuity of electrical displacement indicates that the interior normal field is negligible compared to the exterior normal field. We therefore make the approximation 
\begin{equation}\label{charge approx}
	\sigma= \epsilon_0 (\bE\bcdot\bn)_e,
\end{equation}  
the $e$ subscript indicating evaluation on the exterior side of the interface. 

Consider now the potential distribution in the exterior of the ring, specifically at distances from the centerline comparable with the characteristic thickness $b$. In that  neighborhood of the ring boundary, the exterior potential can be approximated as
\begin{equation}\label{ext pot}
	\varphi(\bx)=-\frac{q(\phi)}{2\pi\epsilon_0}\ln \frac{r}{bf(\phi)}+v(\phi).
\end{equation}
Namely, in any given cross-sectional plane, the exterior potential is locally provided by the potential distribution around an infinite, perfectly conducting, cylinder, which coincides with the ring in that plane and is held at a potential $v(\phi)$. In light of \eqref{charge approx}, the  electrostatic surface-charge density at the cylinder boundary is nothing but the polarization surface-charge density $\sigma$, thence the cylinder's net apparent surface charge, per unit length, is $q(\phi)$. Lastly, the absence of $\theta$-dependent solutions in \eqref{ext pot} can be justified based on the circular shape of the boundary, the uniform potential prescribed on that boundary, as well as the anticipation that the potential gradient is negligible in the azimuthal direction and decays away from the ring's centerline.

We shall now derive two relations between $v(\phi)$ and $q(\phi)$. The first is based on Gauss law, which in the absence of free charge reduces to (cf.~\eqref{QS problem})
\begin{equation}\label{Gauss law}
	\oint dA\,\epsilon_0\epsilon \bE\,\bcdot \bn =0,
\end{equation}
where $dA$ is a differential area element, $\bn$ is an outward normal unit vector and the integral is over an arbitrary closed surface. Consider that closed surface to be the boundary of a curved tube, which closely encloses a segment of the ring between the cross-sectional planes $\phi=\phi'$ and  $\phi=\phi'+\Delta\phi$. From \eqref{q def} and \eqref{charge approx}, the contribution to the integral from the part of the tube boundary that lies outside the ring is $aq(\phi') \Delta\phi$. Since the interior azimuthal field is approximately uniform over the cross section, we find that the contribution from the part of the tube boundary that lies inside the ring, namely the end faces, is
\begin{equation}
	\epsilon_0\mathcal{E}\left\{(A\bE\bcdot\be_{\phi})_{\phi'+\Delta\phi}-(A\bE\bcdot\be_{\phi})_{\phi'}\right\},
\end{equation}
where $A$ denotes the cross-sectional area. By taking the limit $\Delta\phi\to0$, using \eqref{v def}, we find the effective Gauss law 
\begin{equation}\label{Gauss eff}
\frac{q}{\epsilon_0}=\frac{\mathcal{E}}{\kappa^2}\frac{d}{d\phi}\left(\bar{A}\frac{dv}{d\phi}\right),
\end{equation}
where we define the scaled cross-sectional area $\bar{A}=A/b^2$. In the present case, where the cross-sectional shape is circular, $\bar{A}=\pi f^2$.

The second relation arises from the need to resolve the logarithmic growth of the exterior potential away from the centerline of the ring. This is done by matching \eqref{ext pot} with an approximation for the exterior potential that holds at distances from the centerline that are comparable to $a$. Thus, on the latter scale, the finite thickness of the ring is indiscernible and the potential is approximated as that of a circular wire of charge line density $q(\phi)$, \textit{viz.},
\begin{equation}\label{outer}
	\varphi(\bx) = \frac{a}{4\pi\epsilon_0}\int_0^{2\pi} d\phi'\,\frac{q(\phi')}{|\bx-\mathbf{y}(\phi')|}.
\end{equation}
We then compare \eqref{ext pot} with the behavior of \eqref{outer} close to the centerline, 
which is derived in Appendix \ref{app:matching}, to derive the matching condition
\begin{equation}\label{cap}
	v(\phi)=\frac{q(\phi)}{2\pi\epsilon_0}\ln\frac{8\kappa}{f(\phi)}+\frac{1}{4\pi\epsilon_0}\int_0^{2\pi}d\phi'\,\frac{q(\phi')-q(\phi)}{2\sin\frac{|\phi-\phi'|}{2}},
\end{equation}
which is an integral capacitance relation between the voltage at a given azimuthal angle $\phi$ and the azimuthal distribution of polarization charge over the entire ring. 

The differential Gauss law \eqref{Gauss eff} and the integral capacitance relation \eqref{cap} together constitute a `reduced eigenvalue problem' for the `reduced eigenvalue' $\mathcal{E}/\kappa^2$ and `reduced eigenfunctions' $v(\phi)$ and $q(\phi)/\epsilon_0$, which are $2\pi$-periodic functions of $\phi$. Like the exact plasmonic eigenvalue problem, this reduced problem is purely geometric; in particular, the geometry enters through the thickness profile $f(\phi)$ and the logarithm of the aspect ratio $\kappa$.  

\subsection{`Logarithmically approximated' solutions to the reduced eigenvalue problem}
Before attempting to solve the reduced eigenvalue problem in its present form, it is insightful to discuss a coarse `logarithmic' approximation of the reduced problem which is based on the formal largeness of $\ln \kappa$ as $\kappa\to\infty$.  
In this approximation, the integral capacitance relation \eqref{cap} reduces to the algebraic relation
\begin{equation}\label{cap local}
	q/\epsilon_0\approx  \frac{2\pi}{\ln \kappa}v.
\end{equation}
Substituting \eqref{cap local} into the Gauss law \eqref{Gauss eff} then gives
\begin{equation}\label{local ev problem}
	-\frac{\mathcal{E}}{\kappa^2}\frac{d}{d\phi}\left(\bar{A}\frac{dv}{d\phi}\right)+\frac{2\pi}{\ln\kappa}v\approx 0,
\end{equation}
to be solved together with the periodic boundary conditions on $v$. 

The simplified model \eqref{local ev problem} clearly implies the  asymptotic scaling 
\begin{equation}\label{scaling}
	\mathcal{E}\simeq 
	\frac{\kappa^2}{\ln \kappa}
	\quad \text{as} \quad \kappa\to\infty,
\end{equation}
which slightly differs from the usually quoted scaling for slender geometries mentioned in the introduction by the logarithmic factor; we note that the same scaling was found in \cite{Ruiz-Schnitzer:19} for the longitudinal modes of slender bodies of revolution. In particular, consider the case of a torus-shaped ring, with $b$ identified as the cross-sectional radius. Then $\bar{A}(\phi)\equiv\pi$ and solving \eqref{local ev problem} yields the approximate eigenvalues
\begin{equation}\label{E log}
	\mathcal{E}\ub{m}\approx -\frac{2\kappa^2}{m^2\ln \kappa}, \quad m=1,2,\ldots,
\end{equation}
with corresponding pairs of independent eigenfunctions
\refstepcounter{equation}
$$
\label{v approx}
	v\ub{m,c} \approx 
	\cos m\phi, \quad v\ub{m,s} \approx 
	\sin m\phi,
	\eqno{(\theequation{\mathrm{a},\!\mathrm{b}})}
$$
the degeneracy being induced by symmetry.  The corresponding polarization-charge eigenfunctions $q\ub{m,c}$ and $q\ub{m,s}$ can readily be deduced from \eqref{cap local}.

The above scheme corresponds to a leading-order approximation in a perturbative expansion in inverse powers of $\ln \kappa $. In particular, the relative error in \eqref{E log} is on the order of $1/\ln \kappa$. In slender-body theory, such logarithmic approximations are often referred to as `local slender-body theory'. This should be contrasted with the non-approximated  reduced eigenvalue problem, which constitutes a `nonlocal slender-body theory,' the term `nonlocal' here referring to the integral capacitance relation \eqref{cap}. It can be shown that the reduced eigenvalue problem effectively sums the logarithmic expansion mentioned above, resulting in solutions which are `algebraically' accurate, i.e., involving relative errors which are asymptotically smaller than some negative power of $\kappa$.  The distinction between local and nonlocal slender-body theory should not be confused with the notion of spatial nonlocality of the metal's dielectric function, which we do not address in this paper. 

\subsection{`Exact' solutions to the reduced eigenvalue problem}\label{sec:exact}
\subsubsection{Diagonalization of the self-interaction integral operator}
We now proceed to derive `exact' solutions to the reduced eigenvalue problem (cf.~\eqref{Gauss eff} and \eqref{cap}), i.e., without exploiting the formal largeness of $\ln\kappa$. An essential step is to note that the integral operator appearing in the capacitance relation \eqref{cap} is diagonalized by the set of Fourier eigenfunctions. Specifically, for any integer $m$,
\begin{equation}
\label{int identity}
\int_0^{2\pi}d\phi'\,\frac{e^{im\phi'}-e^{im\phi}}{2\sin\frac{|\phi'-\phi|}{2}}=\lambda_m e^{im\phi},
\end{equation}
with $\lambda_0=0$ and 
\begin{equation}\label{eigenvalue int identity}
\lambda_m= -4\sum_{k=1}^{|m|}\frac{1}{2k-1} \quad \text{for} \quad m=\pm1,\pm 2,\ldots
\end{equation}
A proof of this result is given in Appendix \ref{app:identity}. 

\subsubsection{Torus-shaped rings} \label{ssec: torus-shaped rings}
We first consider the case of torus-shaped rings, for which $f$ and $\bar{A}$ are independent of $\phi$. Identifying the cross-sectional radius as $b$, we have $f(\phi)\equiv1$ and $\bar{A}(\phi)\equiv\pi$. (Henceforth, we will always adopt this convention when referring to torus-shaped rings.) For constant $\bar{A}$, the same set of Fourier eigenfunctions that diagonalizes the integral operator in the capacitance relation \eqref{cap} as in \eqref{int identity} also diagonalizes the differential operator in the Gauss law \eqref{Gauss eff}, since $d^2e^{im \phi}/d\phi^2 = -m^2e^{im \phi}$. 
Furthermore, for constant $f$, the logarithmic factor multiplying $q(\phi)$ in \eqref{cap} reduces to a constant. Accordingly, it is immediate to deduce the eigenvalues 
\begin{equation}\label{E exact}
	\mathcal{E}\ub{m} =  -\frac{2\kappa^2}{m^2}
	\left(\ln 8 \kappa - 2\sum_{k=1}^m\frac{1}{2k-1}\right)^{-1}, \quad  \text{for} \quad m=1,2,\ldots,
\end{equation}
with associated voltage eigenfunctions
\refstepcounter{equation}
$$
\label{v exact}
	v\ub{m,c} = 
	\cos m\phi, \quad v\ub{m,\,s} = 
	\sin m\phi.
	\eqno{(\theequation{\mathrm{a},\!\mathrm{b}})}
$$

The closed-form expression \eqref{E exact} constitutes an algebraically accurate slender-body approximation for the reduced eigenvalues of a torus-shaped ring. It is asymptotically consistent with, and improves on, the corresponding logarithmically accurate approximation \eqref{E log}. Surprisingly, the algebraically accurate voltage eigenfunctions \eqref{v exact} are of the same form as the logarithmically accurate eigenfunctions \eqref{v approx}. The algebraically accurate relation between the voltage and polarization-charge eigenfunctions, however, reads
	\begin{equation}\label{q exact}
		q\ub{m,c}/\epsilon_0=2\pi \left(\ln8 \kappa  - 2\sum_{k=1}^m\frac{1}{2k-1}\right)^{-1}
		v\ub{m,c},
	\end{equation}
which should be compared with the corresponding logarithmically accurate relation \eqref{cap local}.

\begin{figure}[]
	\centering \includegraphics[scale=0.7]{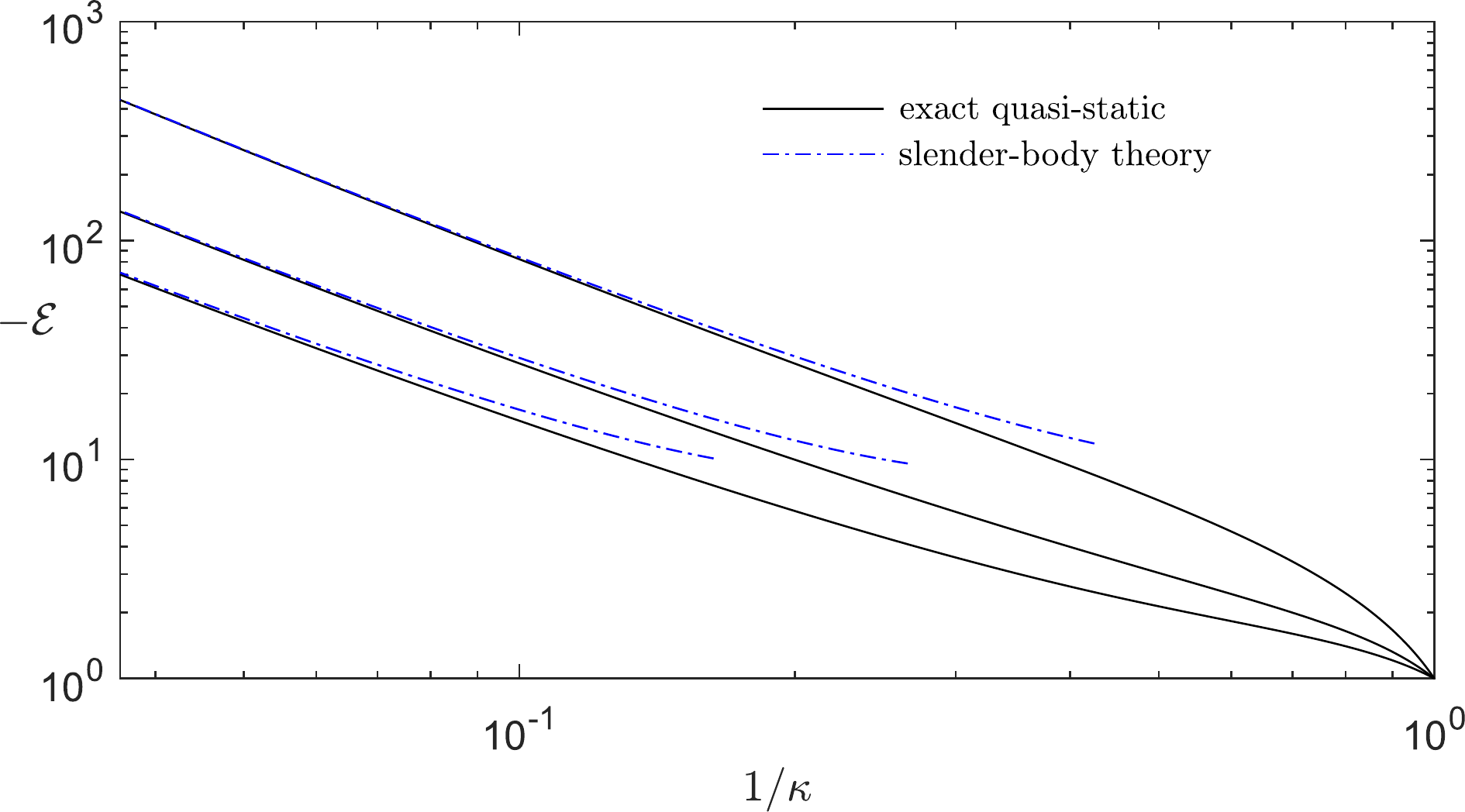}
	\caption{Three lowest permittivity eigenvalues, not counting multiplicities, of a torus-shaped ring as a function of the reciprocal of its aspect ratio $\kappa$: exact solutions of the quasi-static plasmonic eigenvalue problem \cite{Mary:05} vs.~the slender-body approximation \eqref{E exact}.} 
	\label{fig:eig_comparison}
\end{figure}

In the present case of a torus-shaped ring, exact solutions of the plasmonic eigenvalue problem can be obtained by separation of variables in toroidal coordinates \cite{Mary:05}. In Fig.~\ref{fig:eig_comparison}, we compare the first three eigenvalues, not counting multiplicities, computed in this manner with the corresponding values predicted by the slender-body approximation \eqref{E exact}. As expected, the slender-body approximations approach the computed eigenvalues as $\kappa\to\infty$. 

At the same time, the comparison in Fig.~\ref{fig:eig_comparison} demonstrates the fact that, for fixed $\kappa$, the slender-body approximation deteriorates in accuracy with increasing  mode number; conversely, as the mode number increases, accuracy can be retained only by increasing $\kappa$. This is not surprising given that our slender-body approximation corresponds to a limit process where $\kappa\to\infty$ for a fixed mode; there is therefore no reason to expect the approximation to hold in the double limit where both $\kappa$ and $m$ are large \cite{Ruiz-Schnitzer:19}. In fact, the scale on which the voltage and charge eigenfunctions vary along the ring --- assumed large compared to the ring thickness in the slender-body approximation --- becomes comparable to the ring thickness for $m\simeq \kappa$; moreover, for such large $m$, \eqref{E exact} gives $|\mathcal{E}\ub{m}|\simeq 1$, whereas the theory relies on the permittivity contrast being large. We note that this limitation of the theory does not interfere with our goal of describing the low-order (low-frequency) longitudinal resonances of slender nanometallic rings.

The remainder of this section and the next two sections are concerned with extensions of the slender-body theory derived so far to the calculation of the longitudinal modes of more general ring geometries, including azimuthally non-uniform rings, ring dimers and chains and rings with non-circular cross-sections. Some readers may prefer at this point to jump to \S\ref{sec:PW}, where we illustrate the application of the slender-body theory to the solution of the quasi-static scattering problem formulated in \S\ref{sec:QS}.

\subsubsection{Rings of non-uniform thickness}\label{sssec:nonuniform}
Consider now rings for which the thickness profile $f$ depends on the azimuthal angle $\phi$. In this more general case, we solve the reduced eigenvalue problem using a semi-analytical method, which is based on approximating the reduced eigenfunctions $v(\phi)$ and $q(\phi)$ by the truncated Fourier series 
\refstepcounter{equation}
$$
\label{semi_analytical} 
	    v = \sum_{k = 0}^{K}\left\{ \alpha_k \cos k\phi +  \beta_{k} \sin k\phi\right\}, 
		\quad 
		\frac{q}{\epsilon_0} = \sum_{k = 1}^{K}\left\{ \tilde{\alpha}_k \cos k\phi  +  \tilde{\beta}_k \sin k\phi \right\},
		\eqno{(\theequation{\mathrm{a},\!\mathrm{b}})}
$$
where $K\in\mathbb{N}$ is a truncation parameter discussed below. Note that the zeroth harmonic is omitted from the charge representation (\ref{semi_analytical}b) in accordance with the zero-net-charge constraint $\int_{0}^{2\pi}d\phi\,q = 0$, which readily follows from integration of  Gauss law \eqref{Gauss eff}. With \eqref{semi_analytical}, Fourier projection of the capacitance relation \eqref{cap} and Gauss law \eqref{Gauss eff} yields a $2K\times 2K$ generalized eigenvalue problem of the form 
\refstepcounter{equation}
$$\label{numerical scheme}
		\mathsf{q} = 
	-\bar{\mathcal{E}}\,
	\mathsf{M}\boldsymbol{\cdot}\mathsf{v},\quad
			\mathsf{v} = \mathsf{U}\boldsymbol{\cdot}\mathsf{q},
			\eqno{(\theequation{\mathrm{a},\!\mathrm{b}})}
$$
where $\mathsf{v} = (\alpha_1,\ldots,\alpha_K,\beta_1,\ldots,\beta_K)^T$ and $\mathsf{q} = (\tilde{\alpha}_1,\ldots,\tilde{\alpha}_K,\tilde{\beta}_1,\ldots,\tilde{\beta}_K)^T$, with $T$ denoting the vector transpose; we have introduced the notation $\bar{\mathcal{E}}=\mathcal{E}/\kappa^2$ for the reduced eigenvalue; and the matrices $\mathsf{M}$ and $\mathsf{U}$ are provided in Appendix \ref{app:scheme}, along with an expression for the coefficient $\alpha_0$, which is computed \textit{a posteriori}. As already mentioned, the slender-body approximation cannot be expected to correctly capture thickness-scale variations in the longitudinal direction. In fact, the reduced formulation can be shown to be ill posed for $K\simeq \kappa$ \cite{Tornberg:04}. Accordingly, the truncation parameter $K$ should be chosen much smaller than $\kappa$. Of course, $K$ should also be chosen larger than the number of modes to be resolved.

In contrast to the case of a torus-shaped ring, it is clear from the above scheme that for a non-uniform ring each mode generally  contains multiple Fourier harmonics. In particular, whereas for a torus-shaped ring only modes $(1,c)$ and $(1,s)$, which share the degenerate eigenvalue $\mathcal{E}\ub{1}$, are `dipolar', i.e., include a first Fourier harmonic, for non-uniform rings generally all modes include a dipolar component. As we shall see in \S\ref{sec:PW}, this observation is important for interpreting the differences between the resonant  response of uniform and non-uniform rings under plane-wave  illumination.

\begin{figure}[]
	\centering\includegraphics[scale=0.15]{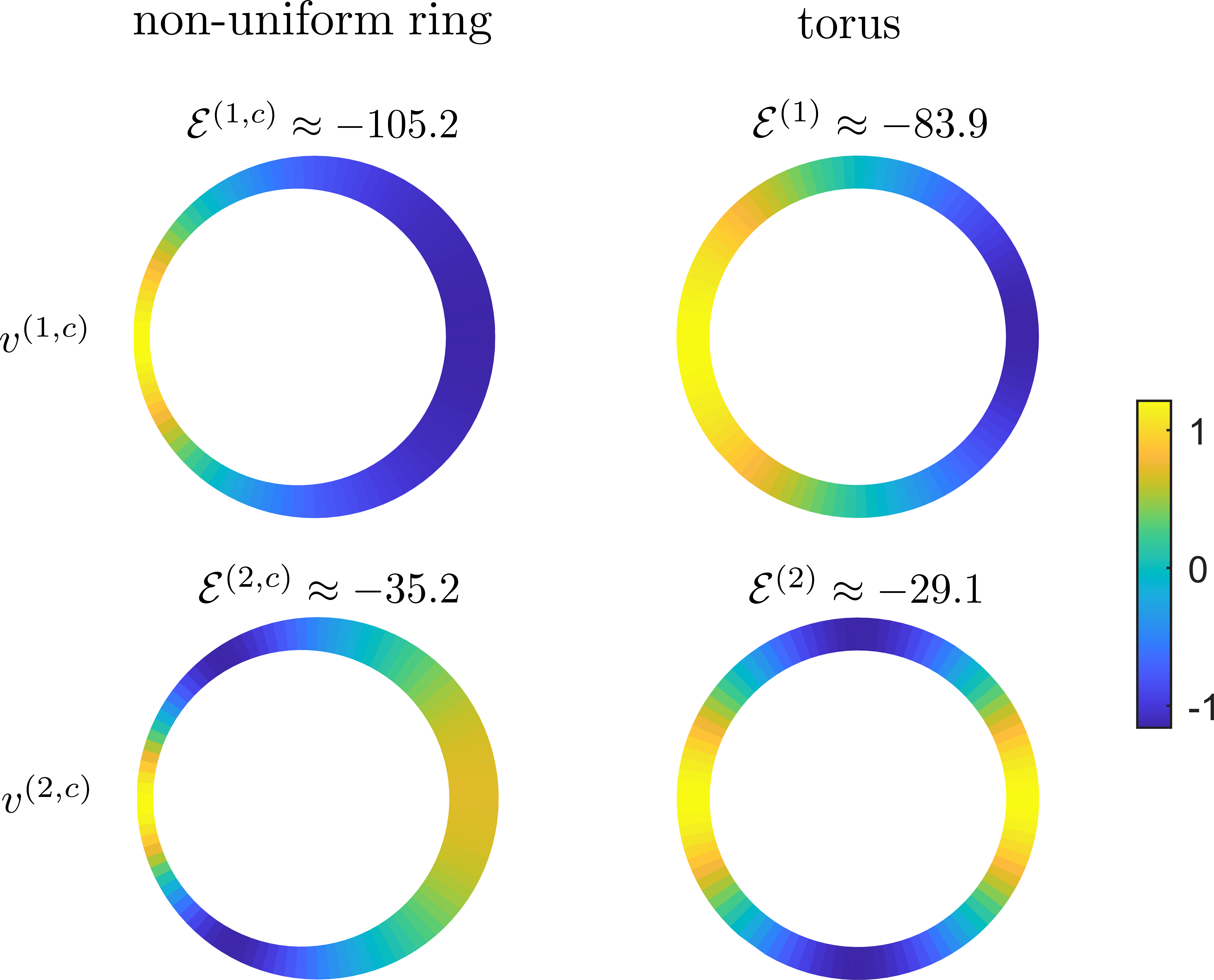}
	\caption{\label{fig:single ring modes comparison}Slender-body approximations for the `dipolar' $(1,c)$ and `quadrupolar' $(2,c)$ voltage eigenfunctions and associated eigenvalues of a torus-shaped ring and an azimuthally non-uniform ring with thickness profile $f(\phi)=1+0.5\cos\phi$; for both geometries $\kappa=10$.} 
\end{figure}

As an example, we consider an azimuthally non-uniform ring whose thickness profile is $f=1+0.5\cos\phi$. The voltage eigenfunctions can be classified as being either even or odd with respect to the symmetry plane of the geometry which is normal to the plane of the ring; referring to (\ref{semi_analytical}a), these modes are respectively comprised of either only cosine or only sine Fourier components. Consistently with our notation for the longitudinal modes of a torus-shaped ring (cf.~\eqref{v exact}), we denote the $m$th even mode by the superscript $(m,c)$; similarly, the $m$th odd mode by the superscript $(m,s)$. In Fig.~\ref{fig:single ring modes comparison}, we show for $\kappa=10$ the `dipolar' $(1,c)$ and `quadrupolar' $(2,c)$ modes of a torus-shaped ring, along with the corresponding modes of the azimuthally non-uniform ring. In the latter case, the modes are computed by solving  \eqref{numerical scheme} with $K = 6$. 

\section{Longitudinal models of ring dimers}\label{sec:dimer}
In this section we consider the longitudinal modes of ring dimers. In \S\ref{ssec:reduced2}, we generalize the reduced eigenvalue problem derived in \S\ref{sec:reduced evp} for a single ring to the dimer case. In \S\ref{sec:coaxial dimer}, we obtain exact closed-form solutions to the generalized reduced problem in the case of a dimer formed of coaxial, generally dissimilar, torus-shaped rings. In \S\ref{sec:generic dimer}, we consider more general ring dimers using a semi-analytical scheme (we also present an \textit{ad hoc} approximation which is in some cases suitable). As we will demonstrate in \S\ref{sec:extensions}, configurations of more than two rings can be handled similarly.
 
\subsection{Reduced eigenvalue problem for ring dimers}\label{ssec:reduced2}
We adopt the same notation as in \S\ref{sec:long}, only with subscripts added to indicate to which ring a given quantity is associated with. By revisiting the derivation in \S\ref{sec:long}, we find that in the thickness-scale vicinity of the rings we have the local approximation
\begin{equation}\label{ext pot2}
	\varphi(\bx)=-\frac{q_n(\phi)}{2\pi\epsilon_0}\ln \frac{r}{b_nf_n(\phi)}+v_n(\phi), \quad n=1,2,
\end{equation}
for the exterior potential, 
which generalizes \eqref{ext pot}. Similarly, the effective Gauss law \eqref{Gauss eff} now applies to each ring separately:
\begin{equation}\label{Gauss coupled}
\frac{q_n}{\epsilon_0} = \frac{\mathcal{E}}{\kappa_n^2}\frac{d}{d\phi}\left(\bar{A}_n\frac{d v_n}{d\phi}\right), \quad n=1,2. 
\end{equation}

It remains to generalize the integral capacitance relation \eqref{cap}. To this end, consider the ring-scale exterior potential  \eqref{outer}, which generalizes as 
\begin{equation}\label{outer2}
	\varphi(\bx) = \sum_{n=1,2}\frac{a_n}{4\pi\epsilon_0}\int_0^{2\pi} d\phi'\,\frac{q_n(\phi')}{|\bx-\mathbf{y}_n(\phi')|}.
\end{equation}
Using the results of Appendix \ref{app:matching} to match \eqref{outer2} and \eqref{ext pot2}, we find the coupled pair of integral capacitance relations
\begin{multline}\label{cap coupled}
v_n(\phi) = \frac{q_n(\phi)}{2\pi \epsilon_0}\ln{\frac{8\kappa_n}{f_n(\phi)}} + \frac{1}{4\pi \epsilon_0}\int_{0}^{2\pi}d\phi'\, \frac{q_n(\phi')-q_n(\phi)}{2\sin\frac{\left|\phi'-\phi\right|}{2}} \\ +\frac{a_k}{4\pi \epsilon_0}\int_{0}^{2 \pi}d\phi'\,\frac{q_k(\phi')}{|\mathbf{y}_n(\phi)-\mathbf{y}_k(\phi')|}, \end{multline}
for $(n,k)=(1,2)$ and $(2,1)$.

The differential equations \eqref{Gauss coupled} and integral equations \eqref{cap coupled} together constitute a generalized reduced eigienvalue problem for ring dimers. We see that the rings interact solely through the last `coupling' integral on the right-hand side of \eqref{cap coupled}, whereby the polarization-charge distribution of one ring induces a voltage disturbance in the other ring, and vice versa. 

\subsection{Coaxial rings}\label{sec:coaxial dimer}
\subsubsection{Diagonalization of the coupling integral operator}
For coaxial rings, the coupling-integral operator in \eqref{cap coupled} is diagonalized by the same Fourier basis that diagonalizes the self-interaction integral operator as in \eqref{int identity}. Indeed, in this case, the distance  between the point on the centerline of ring $1$ at azimuthal angle $\phi_1$ and the point on the centerline of ring $2$ at azimuthal angle $\phi_2$ can be written $|\mathbf{y}_1(\phi_1)-\mathbf{y}_2(\phi_2)|=D(\phi_1-\phi_2)$, where $D(u) = \sqrt{h^2 + a_1^2 + a_2^2 - 2a_1a_2\cos u}$, 
$h$ being  the vertical distance between the rings.
Since $D(u)$ is even, periodic and positive, its reciprocal can be expanded as a cosine Fourier series. It then readily follows that
\begin{equation}
\label{dimer spectral decomposition}
			\int_{0}^{2 \pi}d\phi'\,\frac{e^{i m\phi'}}{|\mathbf{y}_1(\phi)-\mathbf{y}_2(\phi')|} = \tau_me^{i m\phi}, \quad \text{for} \quad m=0,\pm 1,\pm 2,\ldots
\end{equation}
where
\begin{equation}\label{tau def}
\tau_m = \int_{0}^{2 \pi}du\,\frac{\cos mu}{D(u)}.
\end{equation}
\begin{figure}[b!]
	\begin{center}
		\includegraphics[scale=0.45]{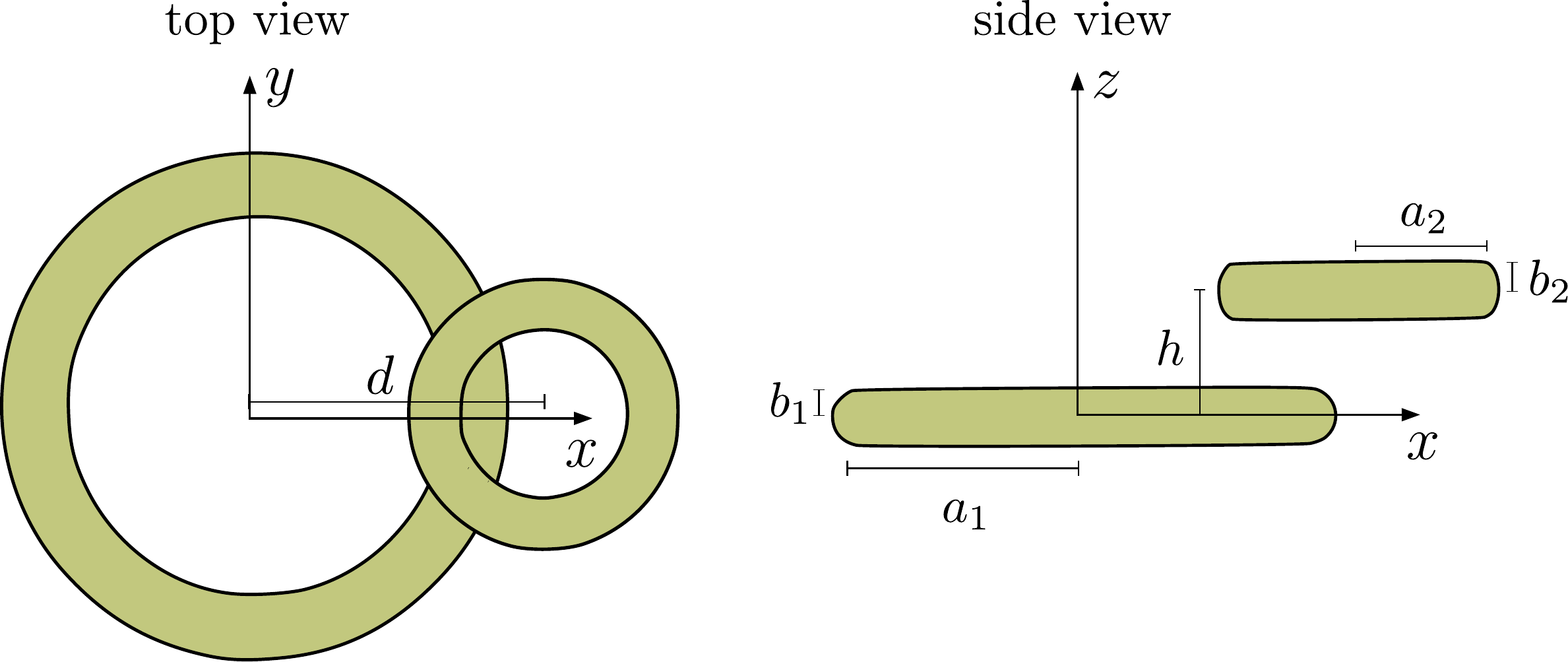}
		\caption{Schematic of a bilayer dimer of torus-shaped rings. The rings have centerline radii $a_1$ and $a_2$, and aspect ratios $\kappa_1 = a_1/b_1$ and $\kappa_2 = a_2/b_2$. The horizontal and vertical separation between the rings is $d$ and $h$, respectively. The coaxial case considered in  \S\ref{sec:coaxial dimer} corresponds to the case $d=0$. The non-coaxial case $d\ne0$ is considered in \S\ref{sec:generic dimer}.}
		\label{fig:schematic_coaxial}
	\end{center}
\end{figure}

With \eqref{dimer spectral decomposition}, it is straightforward to generalize the semi-analytical scheme of \S\ref{sssec:nonuniform} to the case of arbitrary coaxial ring dimers. Instead, we shall focus in the present subsection on the case of coaxial dimers formed of torus-shaped, not necessarily identical, rings (see Fig.~\ref{fig:schematic_coaxial} for $d=0$), where \eqref{dimer spectral decomposition} actually facilitates the derivation of closed-form solutions. Later, in \S\ref{sec:generic dimer}, we will present a more general semi-analytical scheme, not relying on \eqref{dimer spectral decomposition}, that applies to arbitrary ring dimers, including non-coaxial dimers formed of non-uniform rings.  

\subsubsection{Coaxial homodimers}\label{ssec:torusdimersHOMO}
Consider first the case of a coaxial homodimer, namely a coaxial pair of identical torus-shaped rings ($a_1 = a_2 = a$,  $\kappa_1 = \kappa_2 =\kappa$). Given the symmetries of the geometry, we anticipate eigenvalues $\mathcal{E}\ub{m,\pm}$, for $m=1,2,\ldots$, with corresponding eigenfunctions
\refstepcounter{equation}
$$\label{homodimer form} 
	\begin{pmatrix}
		v_{1}\ub{m,c,\pm}\\
		v_{2}\ub{m,c,\pm}
	\end{pmatrix} = \begin{pmatrix}
		1\\
		\pm 1
	\end{pmatrix}\cos m\phi,\quad \begin{pmatrix}
		v_{1}\ub{m,s,\pm}\\
		v_{2}\ub{m,s,\pm}
	\end{pmatrix} = \begin{pmatrix}
		1\\
		\pm 1 
	\end{pmatrix}\sin m\phi.\eqno{(\theequation{\mathrm{a},\!\mathrm{b}})}
$$
The $\pm$ modes are, respectively, even and and odd about the plane equidistance between the rings; we shall also refer to these as in- and out-of-phase modes, respectively, as this will allow a generalized interpretation in the heterodimers case considered next. Writing the eigenfunctions as in \eqref{homodimer form}, use of the diagonalization identities \eqref{int identity} and \eqref{dimer spectral decomposition} readily yields
\begin{equation}
	\label{eigenvalue dimer same size}
	\mathcal{E}\ub{m,\pm} = -\frac{2 \kappa^2}{m^2}\left(\ln 8 \kappa - 2\sum_{k=1}^{m}\frac{1}{2k-1} \pm \frac{\Delta_m}{2} \right)^{-1},
\end{equation}
where $\Delta_m=a\tau_m$ is a dimensionless function of the ratio $h'=h/a$ defined by the quadrature  (cf.~\eqref{tau def})
\begin{equation}\label{DeltaM}
\Delta_m(h')=\int_0^{2\pi}\,du\frac{\cos mu}{\sqrt{h'^2+2-2\cos u}},
\end{equation}
which determines the eigenvalue splitting induced by the interaction between the rings (cf.~\eqref{E exact}). 

The functions $\Delta_m(h')$ are positive. Hence, for any $m$, the in-phase modes are higher energy (less negative permittivity), as one would expect. Furthermore, the functions $\Delta_m(h')$ are monotonically decreasing, asymptotically like $\Delta_m=O(1/h^{2m+1})$ as $h'\to\infty$; this represents the approach to the eigenvalues \eqref{E exact} of a single torus-shaped ring. We also note that $\Delta_m(h')$ is logarithmically singular as $h'\to0$, which is acceptable given that the theory only holds for $h\gg b$, i.e., $h'\gg1/\kappa$. In particular, for $m=1$ we find from \eqref{DeltaM} the behaviours
\refstepcounter{equation}
$$
\label{Delta1 limits}
\Delta_1\sim -2\ln h'+6\ln 2-4 \quad \text{as} \quad h'\to0, \quad \Delta_1\sim \frac{\pi}{h'^3} \quad \text{as} \quad h'\to\infty.
\eqno{(\theequation{\mathrm{a},\!\mathrm{b}})}
$$

In Fig.~\ref{fig:co-axial dimer modes}, the eigenvalues $\mathcal{E}\ub{1,\pm}$ of the in- and out-of-phase dipolar modes, calculated using \eqref{eigenvalue dimer same size}, are depicted by the solid curves as a function of $h/a$, for $\kappa=10$. Also shown are the asymptotic behaviors of the eigenvalues for small and large $h/a$, which follow from the behaviors \eqref{Delta1 limits}.

\subsubsection{Coaxial heterodimers}\label{ssec:torusdimersHETRO}
The above results are easily generalized to allow for dissimilar torus-shaped rings. Since the geometry remains azimuthally invariant, we look for voltage eigenfunctions of the form $\{v_1,v_2\} = \{c_1,c_2\}\times \cos m\phi$, in which $c_1$ and $c_2$ are constant prefactors and $m=1,2,\ldots$ There are also $\pi/2$-rotations of these modes having the same eigenvalues.  
Using the diagonalization identities \eqref{int identity} and \eqref{dimer spectral decomposition}, we find that, for $m=1,2,\ldots$, the reduced eigenvalue problem of \S\ref{ssec:reduced2} is transformed into the $2\times 2$ matrix eigenvalue problem 
\begin{equation}\label{matrix eigenvalue problem}
	\begin{pmatrix}
		c_{1}\\
		c_{2} 
	\end{pmatrix} = 
	-\mathcal{E} 
	\begin{pmatrix}
		\gamma_{1,1} & \gamma_{1,2} \\
		\gamma_{2,1} & \gamma_{2,2}
	\end{pmatrix}\begin{pmatrix}
		c_{1}\\
		c_{2} 
	\end{pmatrix},
\end{equation}
in which the matrix on the right-hand side has the diagonal and non-diagonal elements
\refstepcounter{equation}
$$
\label{matrix entries}
		\gamma_{n,n} = \frac{m^2}{2\kappa_n^2}\left(\ln 8\kappa_n - 2\sum_{k=1}^{m}\frac{1}{2k-1}\right), \quad  
		\gamma_{n,k} = \frac{m^2}{4\kappa_k^2}(a_k\tau_m),
\eqno{(\theequation{\mathrm{a},\!\mathrm{b}})}
$$
respectively, where the products $a_k\tau_m$ are dimensionless geometric factors which determine the coupling between the rings (cf.~\eqref{tau def}). 
Solving the above system gives the eigenvalues
\begin{gather}\label{eigenvalue co-axial dimer}
	\mathcal{E}\ub{m,\pm} = -\frac{2}{\gamma_{1,1}+\gamma_{2,2} \pm \sqrt{(\gamma_{1,1}-\gamma_{2,2})^2 +  4\gamma_{1,2}\gamma_{2,1}}} 
\end{gather}
and corresponding eigenvectors
\begin{equation}\label{c1s}
	\begin{pmatrix}
		c_{1}\ub{m,\pm}\\
		c_{2}\ub{m,\pm}
	\end{pmatrix} = \begin{pmatrix}
		\gamma_{1,2}\\
		\frac{\gamma_{2,2}-\gamma_{1,1} \pm \sqrt{(\gamma_{1,1}-\gamma_{2,2})^2 +  4\gamma_{1,2}\gamma_{2,1}}}{2} 
	\end{pmatrix}.
\end{equation}

In summary, for $m=1,2,\ldots$, there are two degenerate plasmonic eigenvalues $\mathcal{E}\ub{m,\pm}$, given by \eqref{eigenvalue co-axial dimer}, with corresponding eigenfunctions
\refstepcounter{equation}
$$\label{eigenvector cosin}
	\begin{pmatrix}
		v_{1}\ub{m,c,\pm}\\
		v_{2}\ub{m,c,\pm}
	\end{pmatrix} = \begin{pmatrix}
		c_{1}\ub{m,\pm}\\
		c_{2}\ub{m,\pm}
	\end{pmatrix}\cos m\phi,\quad \begin{pmatrix}
		v_{1}\ub{m,s,\pm}\\
		v_{2}\ub{m,s,\pm}
	\end{pmatrix} = \begin{pmatrix}
		c_{1}\ub{m,\pm}\\
		c_{2}\ub{m,\pm}
	\end{pmatrix}\sin m \phi, \eqno{(\theequation{\mathrm{a},\!\mathrm{b}})}
$$
where $c_1\ub{m,\pm}$ and $c_2\ub{m,\pm}$ are given by \eqref{c1s}. 

It can be verified from the above results that $\mathcal{E}\ub{m,+}>\mathcal{E}\ub{m,-}$, namely that for any given $m$ the $+$ mode is higher energy than the $-$ mode. Furthermore, $c_1\ub{m,\pm}$ is positive, whereas $c_2\ub{m, +}$ and $c_2\ub{m, -}$ are positive and negative, respectively. The latter observation suggests referring to the $\pm$ modes as in- and out-of-phase modes, respectively, consistently with the coaxial homodimer case. In contrast to the latter case, where the in- and out-of-phase labeling is associated with a mirror symmetry, here the voltage profile in one ring is not simply the same as or negative of that in the other. We shall see in \S\ref{sec:PW} that this distinction is important for interpreting the resonant response of coaxial dimers. 

\begin{figure}[t!]
	\centering\includegraphics[scale=0.8]{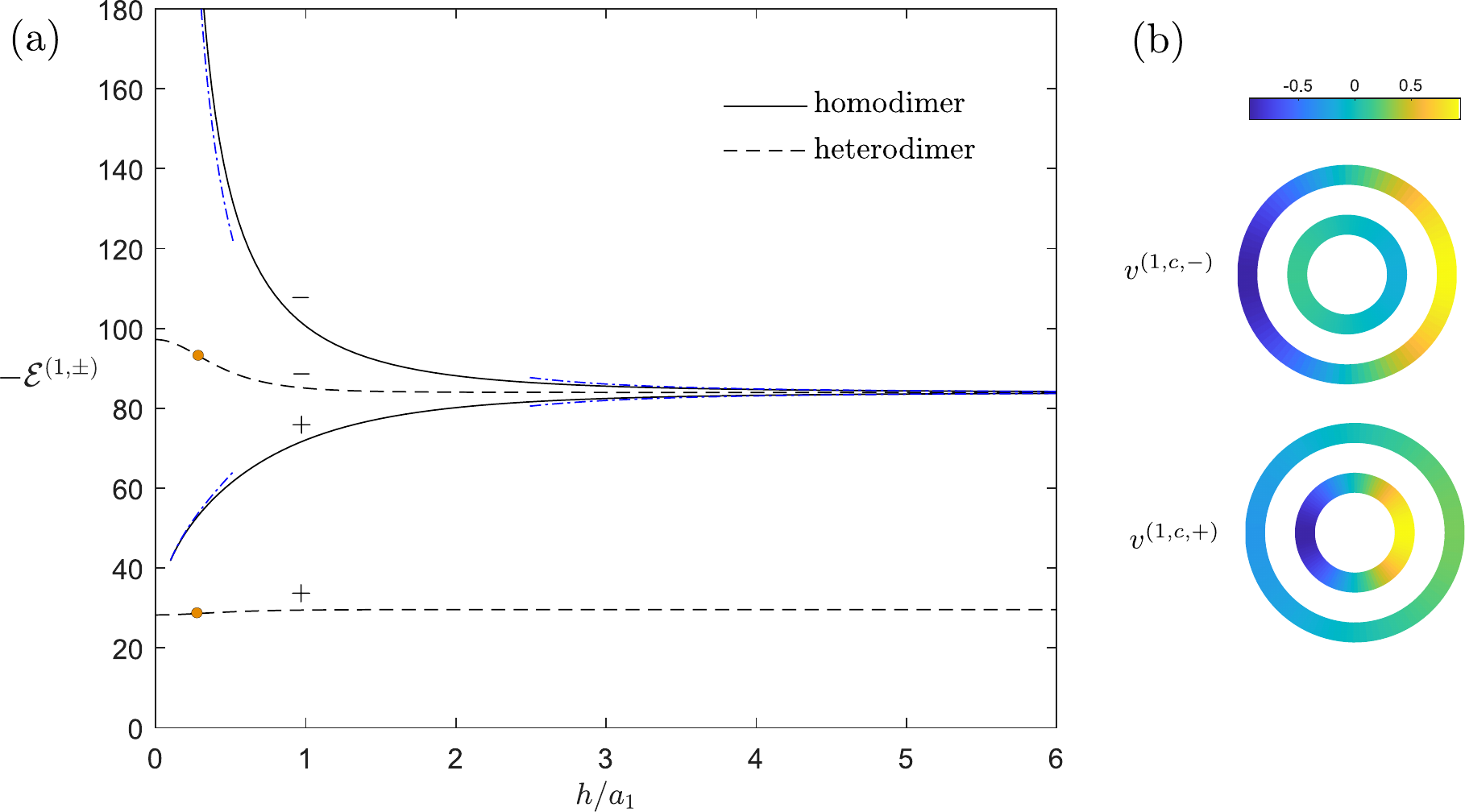}
	\caption{\label{fig:co-axial dimer modes} (a) Eigenvalues of in- and out-of-phase dipolar modes of a a coaxial homodimer ($\kappa=10$) and heterodimer ($\kappa_1=10$, $\kappa_2=5$, $a_1/a_2=2$) as a function of the scaled vertical separation $h/a_1$. The eigenvalues are calculated based on formulas \eqref{E exact} and \eqref{eigenvalue co-axial dimer} for the homodimer and heterodimer, respectively. The dash-dotted lines depict the asymptotic behaviors implied by  \eqref{Delta1 limits}. (b) Corresponding cosine-dipolar voltage eigenfunctions in the heterodimer case, for $h/a_1=0.3$.} \end{figure}

As an example, consider a coaxial heterodimer with $a_1/a_2=2$ and $b_1=b_2$ (i.e., $\kappa_1=10$ and $\kappa_2=5$). In Fig.~\ref{fig:co-axial dimer modes}a, the dashed curves depict the eigenvalues $\mathcal{E}\ub{1,\pm}$, calculated using \eqref{eigenvalue co-axial dimer}, of the in- and out-of-phase dipolar modes of this coaxial heterodimer as a function of $h/a_1$. The voltage profiles for the corresponding cosine-dipolar modes, for $h/a_1=0.3$, are shown in Fig.~\ref{fig:co-axial dimer modes}b. Unlike in the coaxial-homodimer case also presented in that figure, the rings of the heterodimer do not become arbitrarily close as $h\to0$, hence the eigenvalues of the heterodimer approach finite values in that limit. Of course, we could have also considered a coaxial heterodimer formed of rings of similar radius yet different thickness, in which case the eigenvalues would be singular as $h\to0$, as in the coaxial-homodimer case. 

\subsection{Non-coaxial dimers}\label{sec:generic dimer}
\subsubsection{Semi-analytical scheme}\label{sssec:generalscheme}
We now present a semi-analytical scheme to solve the reduced eigenvalue problem of \S\ref{ssec:reduced2} in the case of an arbitrary ring dimer, meaning that the rings need not be coaxial, identical nor azimuthally uniform. Following the derivation of the single-ring semi-analytical scheme in \S\ref{sssec:nonuniform}, we represent the voltage and polarization-charge distributions in each ring by truncated Fourier series as in \eqref{semi_analytical}, with the Fourier coefficients now denoted $\alpha_{n,k},\beta_{n,k}$, etc., with the first and second subscripts corresponding to the ring number and Fourier harmonic, respectively. With that representation, projection of the reduced eigenvalue problem on the truncated Fourier basis yields the $4K\times 4K$ generalized matrix-eigenvalue problem
\refstepcounter{equation}
$$\label{numerical scheme app dimer}
		\begin{pmatrix}
			\mathsf{q}_1\\
			\mathsf{q}_2
		\end{pmatrix} = 
		-{\bar{\mathcal{E}}}
		\begin{pmatrix}
			\dfrac{\kappa_2}{\kappa_1}\mathsf{M}_{1} & 0\\
			0 & \dfrac{\kappa_1}{\kappa_2}\mathsf{M}_{2}
		\end{pmatrix}\boldsymbol{\cdot}\begin{pmatrix}
			\mathsf{v}_1\\
			\mathsf{v}_2
		\end{pmatrix},\quad
		\begin{pmatrix}
			\mathsf{v}_1\\
			\mathsf{v}_2
		\end{pmatrix} = 		\begin{pmatrix}
		\mathsf{U}_{1} & \sqrt{\frac{a_2}{a_1}}\mathsf{V}\\
		\sqrt{\frac{a_1}{a_2}}\mathsf{V}^T & \mathsf{U}_{2}
	\end{pmatrix}\boldsymbol{\cdot}\begin{pmatrix}
			\mathsf{q}_1\\
			\mathsf{q}_2
		\end{pmatrix},
	\eqno{(\theequation{\mathrm{a},\!\mathrm{b}})}
$$
where $\mathsf{v}_n = (\alpha_{n,1},\ldots,\alpha_{n,K},\beta_{n,1},\ldots,\beta_{n,K})^T$ and $\mathsf{q}_n = (\tilde{\alpha}_{n,1},\ldots,\tilde{\alpha}_{n,K},\tilde{\beta}_{n,1},\ldots,\tilde{\beta}_{n,K})^T$; $\bar{\mathcal{E}}=\mathcal{E}/(\kappa_1\kappa_2)$ is a rescaled eigenvalue;  $\mathsf{M}_n$ and $\mathsf{U}_n$ are  identical to the matrices $\mathsf{M}$ and $\mathsf{U}$, respectively, used in the single-ring scheme \eqref{numerical scheme}, with the geometric parameters being those of ring $n$; and $\mathsf{V}$ is a coupling matrix whose  form is provided in Appendix \S\ref{app:scheme dimer}. 

\subsubsection{Non-coaxial homodimers}
\begin{figure}[]	\centering\includegraphics[scale=0.65]{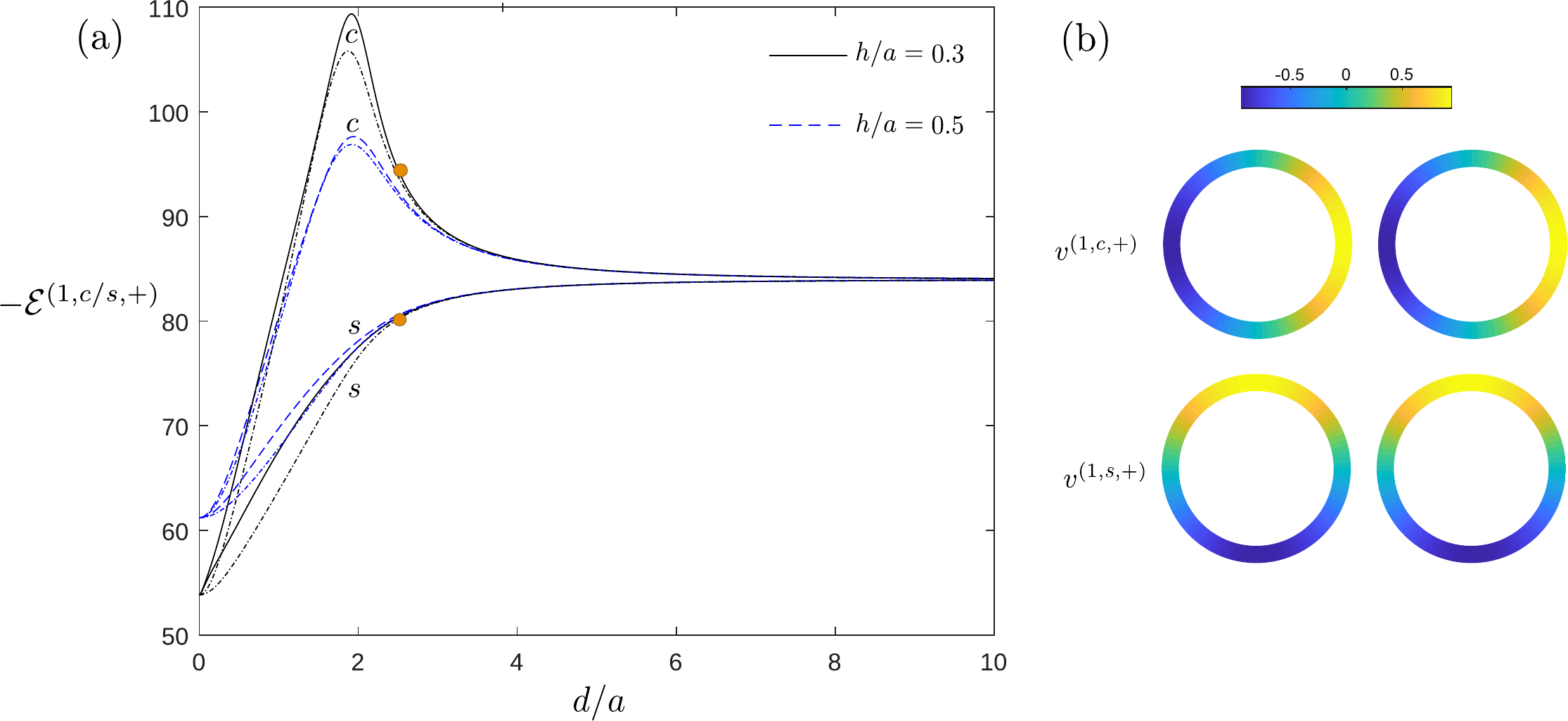}	\caption{\label{fig:homogeneous dimer eigenvalue evol.}(a) Eigenvalues $\mathcal{E}\ub{1,c,+}$ and $\mathcal{E}\ub{1,s,+}$ of the non-coaxial bilayer dimer shown in Fig.~\ref{fig:schematic_coaxial}, in the case where the rings are identical ($\kappa=10$). The eigenvalues, obtained by solving the semi-analytical scheme \eqref{numerical scheme app dimer}, are plotted as a function of the scaled horizontal displacement $d/a$ and for the indicated values of the relative vertical displacement $h/a$.
Dash-dotted curves depict the \textit{ad hoc} analytical approximation \eqref{eigenvalue perturbation}. 
   (b) Corresponding cosine-dipolar and sine-dipolar voltage eigenfunctions in the non-coaxial bilayer dimer for $d/a=2.5$ and $h/a=0.3$.
}
\end{figure}

As an example of a non-coaxial dimer configuration, we consider the bilayer configuration shown in Fig.~\ref{fig:schematic_coaxial}. It consists of a pair of torus-shaped rings whose centerlines define parallel planes separated by the vertical distance $h$, and whose symmetry axes are separated by the horizontal distance $d$. The case where $d$ vanishes corresponds to the coaxial configuration considered in \S\ref{sec:coaxial dimer}, while the limit $h^2+d^2\to\infty$ corresponds to that of non-interacting rings. We have seen that both of these extreme cases are analytically solvable, with each longitudinal mode involving only a single Fourier harmonic. In contrast, in the general case each mode is expected to consist of a combination of Fourier harmonics. Symmetry still allows, however, the modes of the bilayer geometry to be classified based on whether the voltage eigenfunctions are even or odd about the $x$--$z$ mirror plane. With the azimuthal angle $\phi$ measured from the $x$ direction, these even and odd modes involve only cosine or sine Fourier harmonics, respectively. Thus, the semi-analytical scheme \eqref{numerical scheme app dimer} is reduced in this case to two uncoupled $2K\times2K$ matrix problems. 

We first consider the homodimer case where the rings are identical ($a_1=a_2=a$, $\kappa_1=\kappa_2=\kappa$). In particular, we focus attention on the two modes continuated, as $d$ is increased from zero, from the degenerate in-phase dipolar modes $(1,c,+)$ and $(1,s,+)$ found in the coaxial case (\S\ref{ssec:torusdimersHOMO}). Fig.~\ref{fig:schematic_coaxial} depicts the variation with $d/a$ of the respective eigenvalues, say $\mathcal{E}\ub{1,c,+}$ and $\mathcal{E}\ub{1,s,+}$, calculated using the semi-analytical scheme \eqref{numerical scheme app dimer} for two values of $h/a$. Note that the `$c$' and `$s$' modes are even and odd about the $x$--$z$ plane, respectively, as in the coaxial case, but are no longer degenerate rotations of each other. The insets show the voltage eigenfunctions for the indicated values of $d/a$ and $h/a$. 

A numerical study based on the semi-analytical scheme \eqref{numerical scheme app dimer} suggests that the modes of this bilayer homodimer configuration are, by visual inspection,   dominated by a single Fourier harmonic. This harmonic corresponds to that of the mode of the corresponding coaxial (or isolated-ring) configuration from which the bilayer mode is continuated from. This observation suggests an  intuitive and \textit{ad hoc} approximation in which the voltage profiles in both rings are constructed from just the apparently  dominant harmonic. With that assumption, we obtain
\begin{equation}\label{eigenvalue perturbation}
	\mathcal{E}\ub{m,c,\pm} \approx -\frac{2 \kappa^2}{m^2 }\left(\ln 8 \kappa - \sum_{k=1}^{m}\frac{2}{2k-1}  \pm \frac{a}{2\pi}\int_{0}^{2 \pi}\int_{0}^{2 \pi}d\phi_1d\phi_2\,\frac{\cos m\phi_1\cos m\phi_2}{|\mathbf{y}_1(\phi_1)-\mathbf{y}_2(\phi_2)|} \right)^{-1}
\end{equation}
along with a similar expression for $\mathcal{E}\ub{m,s,\pm}$ where $\cos m\phi_1\cos m\phi_2$ is replaced by $\sin m\phi_1\sin m\phi_2$. In Fig.~\ref{fig:homogeneous dimer eigenvalue evol.}, the dash-dotted curves depict  approximation \eqref{eigenvalue perturbation} for $\mathcal{E}\ub{1,c,+}$ and $\mathcal{E}\ub{1,s,+}$. Despite the approximation being heuristic, the agreement with the semi-analytical slender-body scheme \eqref{numerical scheme app dimer} is reasonably good for all $d/a$ and especially at moderately large $d/a$. It is clear that this approximation is exact for $d=0$ and asymptotically correct as $d/a\to\infty$. 

\subsubsection{Non-coaxial heterodimers}\label{sssec:noncoaxialhetero}
\begin{figure}[]
	\centering\includegraphics[scale=0.8]{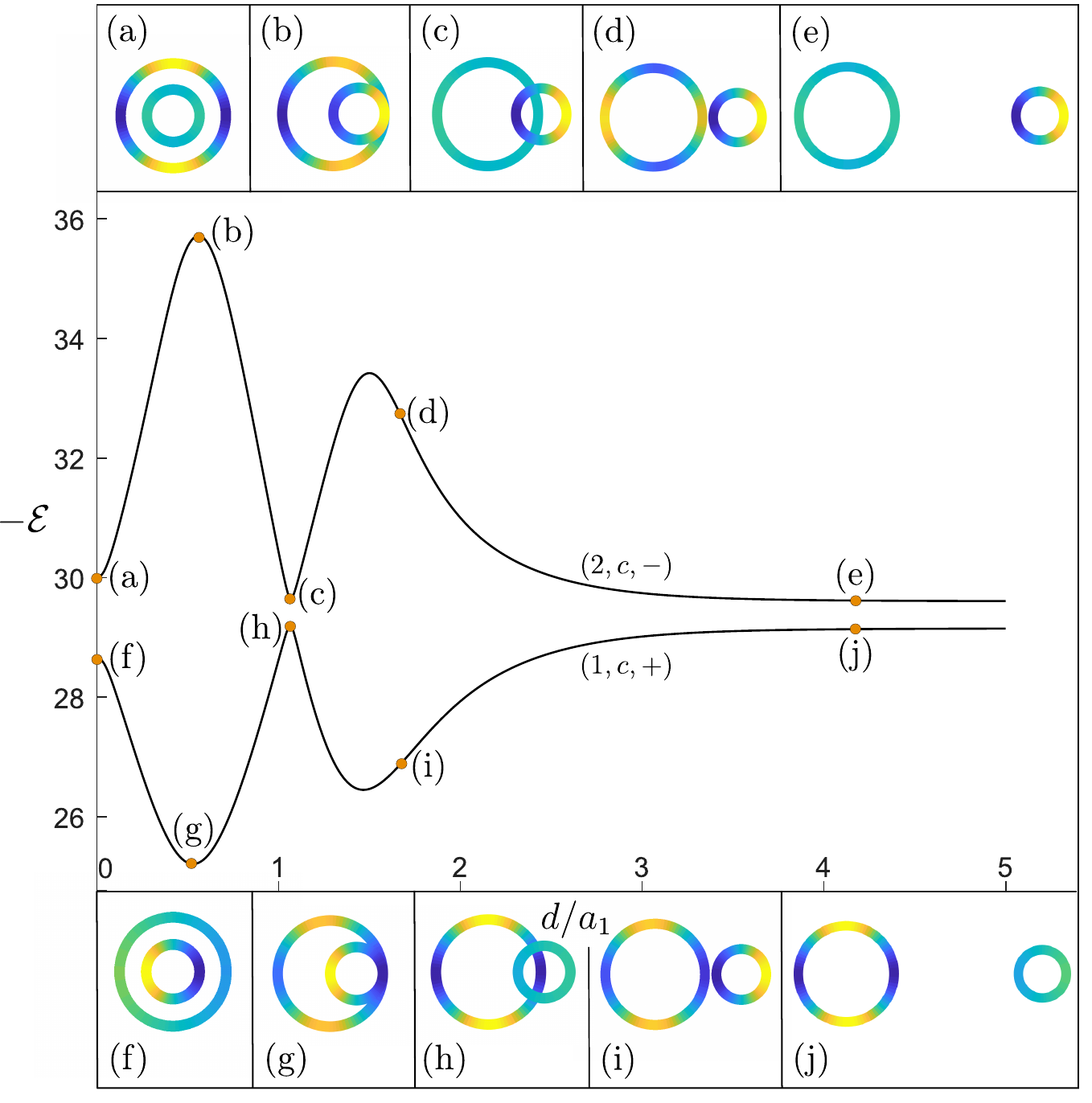}
		\caption{\label{fig:heterogeneous mode evol.} Evolution as a function of $d/a_1$ of two eigenvalues of the bilayer heterodimer configuration schematically shown in Fig.~\ref{fig:schematic_coaxial}, in the case $\kappa_1 = 10$, $\kappa_2=5$, $a_1/a_2 = 2$ and $h/a_1=0.3$. The modes are those continuated from the in-phase dipolar mode $(1,c,+)$ and out-of-phase quadrupolar mode $(2,c,-)$, respectively, of the coaxial configuration obtained for  $d=0$. Also shown are the voltage eigenfunctions at the indicated values of $d/a_1$.}
\end{figure}

Consider now a bilayer heterodimer where the torus-shaped rings are not identical. Unlike in the bilayer homodimer scenario considered above, now the spectrum of Fourier harmonics comprising each mode undergoes significant evolution as the horizontal displacement $d$ is increased from the coaxial case $d=0$, or, alternatively, decreased from the isolated-ring case $d=\infty$. Moreover, we find that this evolution can continuously link two different Fourier harmonics in the latter limiting cases, in which each mode is composed of a  single Fourier harmonic. We demonstrate this in Fig.~\ref{fig:heterogeneous mode evol.} by considering the evolution of the two modes continuated from the coaxial in-phase dipolar mode $({1,c,+})$ and out-of-phase quadrupolar mode $({2,c,-})$, respectively, with the geometric parameters of the two rings chosen such that the eigenvalues are close in the coaxial configuration. In the coaxial configuration, the mode $(2,c,-)$ is dominated by a quadrupolar distribution in the larger ring, whereas the mode $(1,c,+)$ is dominated by a dipolar distribution in the smaller ring. As $d$ is continuously increased to $\infty$, the mode $(2,c,-)$ is ultimately dominated by the dipolar mode $(1,c)$ of the smaller ring; similarly, the mode $(1,c,+)$ is ultimately  dominated by the quadrupolar mode $(2,c)$ of the larger ring. This evolution is seen to involve multiple stages during which the Fourier harmonics of each mode undergo mixing and the eigenvalues split apart and then re-approach each other twice. We will see in \S\ref{sec:PW} some of the implications of this evolution in the context of the scattering problem.

\section{Further geometric extensions}\label{sec:extensions}
\subsection{Arbitrary cross-sectional shapes}\label{ssec:noncirc}
It is straightforward to extend our theoretical framework to allow for non-circular cross-sectional shapes. In particular, let us briefly revisit the derivation in \S\ref{sec:reduced evp} of the single-ring reduced eigenvalue problem. Given our focus on longitudinal modes of slender rings, we still expect that the interior potential is approximately uniform over the ring's cross section and varies mainly in the azimuthal direction. Thus, the representation \eqref{v def} of the interior potential by an azimuthal voltage profile still holds. The exterior potential in the vicinity of the ring, however, can no longer be approximated as in \eqref{ext pot}, since that radially symmetric distribution assumes that the cross sections are circular. Nonetheless, it can be shown that \eqref{ext pot} still holds at intermediate radial distances from the centerline, i.e.,~$b\ll r\ll a$, if only the cross-sectional radius $bf(\phi)$ is  replaced by the so-called `conformal radius', say $bf^*(\phi)$, of the cross-sectional geometry at the azimuthal angle $\phi$. Working in the corresponding cross-sectional plane, the conformal radius $bf^*(\phi)$ can be extracted from a conformal mapping from the exterior of a circle of that radius to the domain exterior to the true cross section (see, e.g., \cite[Chapter~5]{Hinch:91} and \cite{Schnitzer:17}). In particular, for elliptical cross sections with semi-diameters $b\sigma_1(\phi)$ and $b\sigma_2(\phi)$, one finds $bf^*(\phi)=(b\sigma_1(\phi)+b\sigma_2(\phi))/2$. Expressions for several other geometries can be found in \cite[Table~1]{Paquin:20}. As a consequence, $f(\phi)$ should be replaced by $f^*(\phi)$ in the capacitance relation \eqref{cap}, while \eqref{Gauss eff} remains unchanged, with $\bar{A}(\phi)$ still denoting the scaled cross-sectional area $A(\phi)/b^2$. The significance of the present extension is that, in contrast to the case of circular cross sections, $\bar{A}(\phi)$ can now be tuned independently from ${f^*}(\phi)$. The generalization of this extension to the case of ring dimers is evident. 

Consider for example an azimuthally uniform ring whose cross-sectional shape is arbitrary. The eigenvalues follow from the result \eqref{E exact} for a torus-shaped ring as
\begin{equation}\label{E arbitrary cross-section}
	\mathcal{E}\ub{m} =  -\frac{2\pi\kappa^2}{m^2\bar{A}}
	\left(\ln \frac{8 \kappa}{f^*} - 2\sum_{k=1}^m\frac{1}{2k-1}\right)^{-1}, \quad  \text{for} \quad m=1,2,\ldots
\end{equation} 
Note the logarithmically weak influence of the ring's cross-sectional shape, which we emphasize is a specific feature of the longitudinal modes considered herein. 
This prediction is consistent with the experimental and numerical results in \cite{Hao:08}, which show only slight differences between the plasmon-resonance frequencies of rings of different cross-sectional shape yet similar cross-sectional area. 

\subsection{Chain of rings}\label{ssec:chain}
For clarity of exposition, we have so far considered either single rings or ring dimers. It is straightforward, however, to extend our approximation scheme to a system consisting of an arbitrary number of interacting rings. In particular, here we consider a coaxial chain of $N$ not necessarily identical azimuthally invariant rings. (In light of the preceding generalization, azimuthally invariant does not necessarily imply torus-shaped.) In this case,  the appropriate reduced eigenvalue problem can be treated analytically. The analysis closely follows the derivation in \S\S\ref{sec:coaxial dimer} for a coaxial dimer of torus-shaped rings. Similar to the latter case, symmetry implies collective modes, with azimuthal number $m=1,2,\ldots$, in which the voltage eigenfunction in the $n$th ring is $v_n = c_n\cos m\phi$. (There are also $\pi/2$ rotations of these collective modes having the same eigenvalues.) Following the steps in \S\S\ref{sec:coaxial dimer}, for given $m$ we find the $N\times N$ generalized matrix-eigenvalue problem (cf.~\eqref{matrix eigenvalue problem})
\begin{equation}\label{N matrix eigenvalue problem}
	\mathsf{c} = 
	-\mathcal{E}\, 
	\mathsf{G}\boldsymbol{\cdot}\mathsf{c}
\end{equation}
where $\mathsf{c} = (c_1,c_2,\ldots,c_N)^T$ and $\mathsf{G}$   
is a $m$-dependent $N\times N$ matrix whose components are defined  analogously to \eqref{matrix entries}. 
\begin{figure}[p!]
\begin{center}
\includegraphics[scale=0.53]{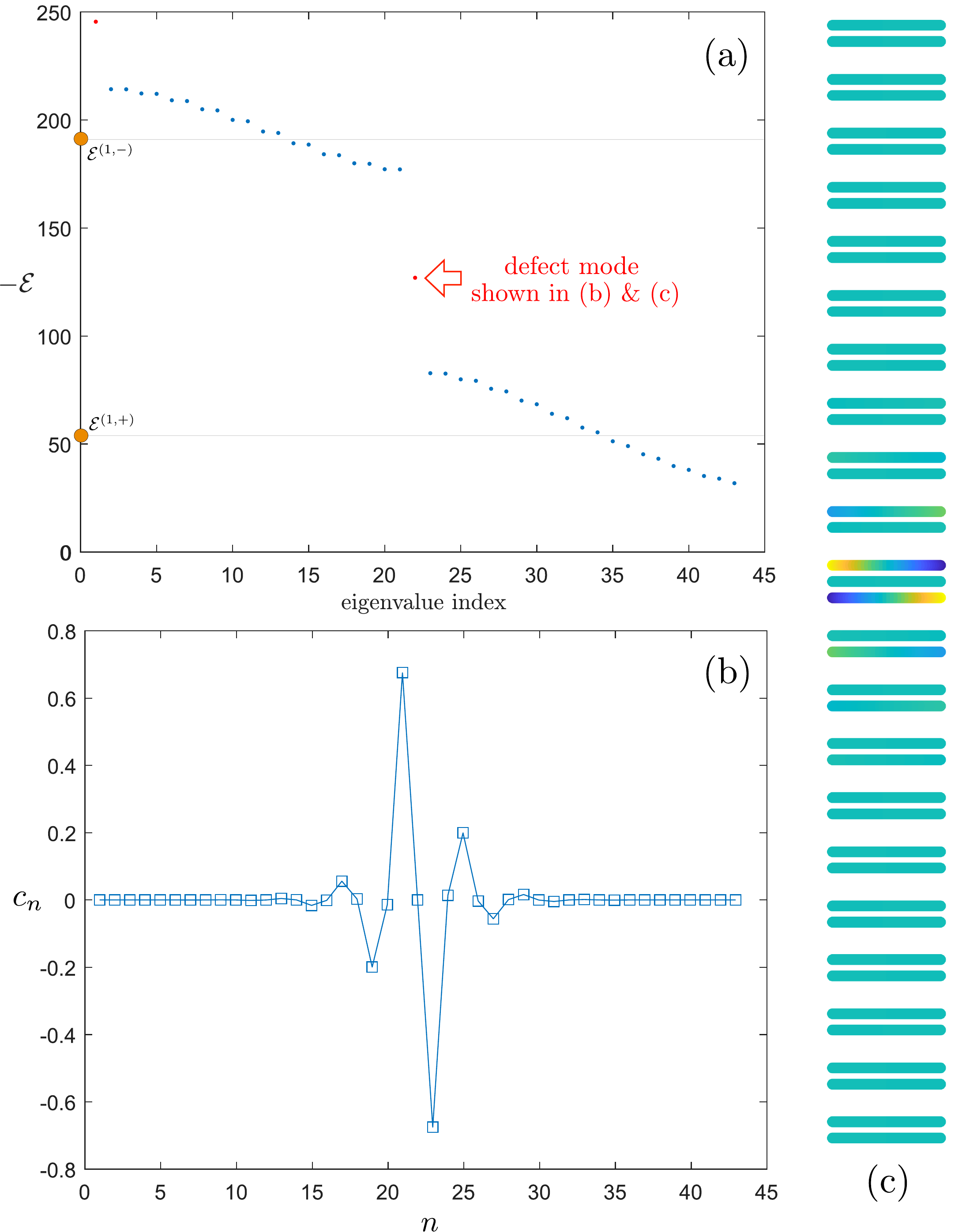}
\caption{(a) Eigenvalues of the `dipolar' ($m=1$) modes of a coaxial chain of $N=43$ identical torus-shaped rings ($\kappa=10$). The chain consists of two identical sub-chains, each formed of $10$ dimers  (spacing between rings and dimers being $0.3$ and $0.7$ times the ring radius, respectively), which are linked by a trimer (spacings similar to dimer sub-chains). Each eigenvalue is doubly degenerate, corresponding to a $\pi/2$ rotation of the eigenfunctions about the symmetry axis. The spectrum is calculated in the slender-body approximation by solving \eqref{N matrix eigenvalue problem}. It consists of two nearly continuous bands that cross the eigenvalues $\mathcal{E}\ub{1,\pm}$ of the dimers in isolation. Additionally, there are two isolated eigenvalues which are associated with localized `defect modes'.  The defect mode marked in (a) is visualized in terms of the dipole intensities $c_n$ (cf.~\eqref{N matrix eigenvalue problem}) in (b) and voltage profiles in (c). The horizontal lines in (a) mark the eigenvalues $\mathcal{E}\ub{1,-}$ and $\mathcal{E}\ub{1,+}$ of the dimers in isolation.}
\label{fig:chain}
\end{center}
\end{figure}

As an example, we employ \eqref{N matrix eigenvalue problem} to calculate the eigenfunctions of a finite coaxial chain of torus-shaped rings. Specifically, we consider a chain consisting of a series of identical, equally spaced, homodimers, where the center dimer is replaced by a homotrimer. In Fig.~\ref{fig:chain}, we focus our attention on the dipolar modes of the chain ($m=1)$, for which the permittivity-eigenvalue spectrum is seen to consist of two disjoint, nearly continuous, bands formed of a large number of densely distributed eigenvalues; the lower- and higher-energy bands respectively cross the eigenvalues $\mathcal{E}\ub{1,-}$ and $\mathcal{E}\ub{1,+}$ of the dimers in isolation (\S\ref{sec:coaxial dimer}). Additionally, there are  two isolated eigenvalues, one in the spectral gap between the two bands and one at an energy lower than the lower-energy band. As demonstrated in the figure, the isolated eigenvalues correspond to modes that are localized around the defect. We remark that \eqref{N matrix eigenvalue problem} predicts localized modes for all $m$, not only for the dipolar modes. In retrospect, the existence of localized modes rationalizes the application of our quasi-static theory to the case of an extended chain of rings, whose total length in any realistic scenario would be at least comparable to the free-space wavelength. Accordingly, we expect non-localized modes of the chain to be significantly affected by retardation \cite{Pocock:2018}.

\section{Plane-wave illumination}
\label{sec:PW} 
Armed with our slender-body approximations for the longitudinal modes of slender-ring structures, we  return to the quasi-static scattering problem formulated in \S\ref{ssec:qs}. There, the near-field potential $\varphi(\bx)$ generically possesses the asymptotic far-field behavior 
\cite{Jackson:07} 
\begin{equation}\label{far field generic}
	\varphi(\mathbf{x}) \sim  -\mathbf{E}_{\infty}\boldsymbol{\cdot}\mathbf{x} +  \mathbf{E}_{\infty}\boldsymbol{\cdot}\boldsymbol{\alpha}\boldsymbol{\cdot}\frac{\mathbf{x}}{4\pi |\mathbf{x}|^3} \quad\text{as}\quad |\mathbf{x}|\to\infty,
\end{equation}
in which $\boldsymbol{\alpha}$ is the polarizability tensor of the structure. In terms of that tensor,  quasi-static approximations for the extinction and absorption cross sections in the direction of the applied field, say $\hat{\textbf{\textit{\i}}}$, are given by \cite{Bohren:Book}
\begin{equation}\label{C def}
C_{\text{ext}},C_{\text{abs}} = \frac{2\pi}{\lambda}\,\hat{\textbf{\textit{\i}}}\hat{\textbf{\textit{\i}}}\boldsymbol{:}\text{Im}\,\boldsymbol{\alpha},
\end{equation}
where $\lambda$ denotes the wavelength of the incident plane-wave. We shall use the approximation scheme developed in this paper to calculate $\boldsymbol{\alpha}$ and hence  $C_{\text{abs}}$ for a range of slender-ring geometries. 

A slender-body approximation for $\boldsymbol{\alpha}$ can be extracted by considering the large $|\bx|$ expansion of the spectral solution \eqref{Field expansion} and substituting the slender-body approximations for the voltage eigenfunctions and eigenvalues. For a structure formed of $N$ arbitrarily shaped rings, we find
\begin{equation}\label{abs}
\boldsymbol{\alpha} =  \frac{1}{\epsilon_0}\sum_{I \in \mathcal{I}}\frac{\epsilon_r(\omega)-1}{\epsilon_r(\omega)-\mathcal{E}^{(I)}}\frac{\left(\sum_{n=1}^N a_n \int_0^{2\pi}d\phi\,\mathbf{y}_nq_n\ub{I}\right)\left(\sum_{n=1}^N a_n \int_0^{2\pi}d\phi\,\mathbf{y}_nq_n\ub{I}\right)}{\sum_{n=1}^{N}a_n\int_0^{2\pi}d\phi\,q_n\ub{I}v_n\ub{I}}.
\end{equation}
A derivation of this result is given in Appendix \ref{appendix far-field}. We note that it only includes the longitudinal modes studied in this paper, which for low frequencies $\omega\ll\omega_p$ dominate the plasmonic response as further discussed in \cite{Ruiz-Schnitzer:19} and the concluding section \S\ref{sec:concluding}.

To evaluate \eqref{abs} for general ring geometries, we employ the semi-analytical scheme \eqref{numerical scheme app dimer}, whose generalization from the case of two to $N$ rings is straightforward; furthermore, non-circular cross sections can be included in that scheme as discussed in \S\ref{ssec:noncirc}. 

In the special cases of single azimuthally invariant rings, as well as azimuthally invariant coaxial dimers, it is possible to evaluate the integrals appearing in \eqref{abs} in closed form. In those cases, symmetry dictates that the polarizability tensor possesses the form $\boldsymbol{\alpha}= (\be_x\be_x+\be_y\be_y)\alpha$, where $\be_x$ and $\be_y$ are orthogonal unit vectors parallel to the plane defined by the rings and $\alpha$ is a scalar polarizability. Furthermore, each mode involves a single Fourier harmonic, and since the centerlines $\mathbf{y}_n$ are circular curves, the overlap integrals in the numerator of \eqref{abs} vanish for all except the dipolar modes. Assuming for simplicity that the azimuthally invariant rings are torus-shaped, we use the results of \S\ref{sec:long} to find
\begin{equation}\label{abs torus}
\alpha= \frac{\epsilon_r(\omega)-1}{\epsilon_r(\omega)-\mathcal{E}^{(1)}} \frac{2\pi^2a^3}{\ln 8\kappa -2},
\end{equation}
with the slender-body approximation for $\mathcal{E}\ub{1}$ provided by \eqref{E exact}.
Similarly, using the results of \S\ref{sec:coaxial dimer}, we find for a coaxial dimer of torus-shaped rings
\begin{equation}\label{abs co-axial dimer}
{\alpha} = \sum_{\pm}\frac{1-\epsilon_r(\omega)}{\epsilon_r(\omega)-\mathcal{E}\ub{1,\pm}}\frac{\pi\mathcal{E}\ub{1,\pm}\left(A_1c_1\ub{1,\pm}+A_2c_2\ub{1,\pm}\right)^2}{\frac{A_1}{a_1}\left(c_1\ub{1,\pm}\right)^2 + \frac{A_2}{a_2}\left(c_2\ub{1,\pm}\right)^2},
\end{equation}
with $c_1\ub{1,\pm}$ and $c_1\ub{2,\pm}$ provided by \eqref{c1s} and a slender-body approximation for   $\mathcal{E}\ub{1,\pm}$ provided by \eqref{eigenvalue co-axial dimer}. In the more specific coaxial homodimer case, the numerator in \eqref{abs co-axial dimer} vanishes identically for the out-of-phase dipolar mode, namely because the induced dipole excited in the two rings cancel. Thus, \eqref{abs co-axial dimer} degenerates to 
\begin{equation}\label{abs homogeneous co-axial dimer}
\alpha= \frac{\epsilon_r(\omega)-1}{\epsilon_r(\omega)-\mathcal{E}^{(1,+)}} \frac{4\pi^2a^3}{\ln 8\kappa -2+\frac{1}{2}\Delta_1},
\end{equation}
where $\mathcal{E}\ub{1,+}$ now possesses the slender-body approximation \eqref{eigenvalue dimer same size} and $\Delta_1$ is given by \eqref{DeltaM}. 

In formulas \eqref{abs}--\eqref{abs homogeneous co-axial dimer}, resonance of mode ${I}$ is associated with  cancellation in the denominator of the frequency-dependent factor $F\ub{I}=(\epsilon_r(\omega)-1)/(\epsilon_r(\omega)-\mathcal{E}\ub{I})$. Let $\mathcal{E}\ub{I}_{0}$ denote the slender-body approximation for $\mathcal{E}\ub{I}$. We stress that the approximation $F\approx (\epsilon_r(\omega)-1)/(\epsilon_r(\omega)-\mathcal{E}\ub{I}_0)$ only holds if $|\mathcal{E}\ub{I}-\mathcal{E}\ub{I}_0|\ll \mathrm{Im}(\epsilon_r(\omega))$, namely if the loss is large relative to the error in the slender-body approximation for the eigenvalue. On the one hand, the algebraic order of the latter error is smaller than $\kappa^2$, though we do not know whether it is unity, $\kappa$ or some other scaling $\ll\kappa^2$. On the other hand, the Drude model \eqref{drude} and eigenvalue scaling \eqref{scaling} together imply $\mathrm{Im}(\epsilon_r(\omega))\simeq \kappa^3(\gamma/\omega_p)$, suggesting that the condition is easily met for the longitudinal resonances. This is confirmed by the comparison below with exact quasi-static solutions. Even if the condition $|\mathcal{E}\ub{I}-\mathcal{E}\ub{I}_0|\ll \mathrm{Im}(\epsilon_r(\omega))$ is not met, formulas \eqref{abs}--\eqref{abs homogeneous co-axial dimer} still give a valid approximation for the  resonance curve only shifted in frequency; moreover, the approximation $F\ub{I}\approx \mathcal{E}\ub{I}_0/\mathrm{Im}(\epsilon_r)$ still holds at resonance, i.e., at $\omega$ such that $\mathrm{Im}(\epsilon_r)=\mathcal{E}\ub{I}$.

We now apply the above results for the scattering problem to several ring structures whose plasmonic eigenvalues and eigenfunctions we have analyzed earlier in the paper. In particular, we consider single torus-shaped and azimuthally non-uniform rings, as well as coaxial and non-coaxial dimers of torus-shaped rings. 
 
\begin{figure}[]
	\centering\includegraphics[scale=0.6]{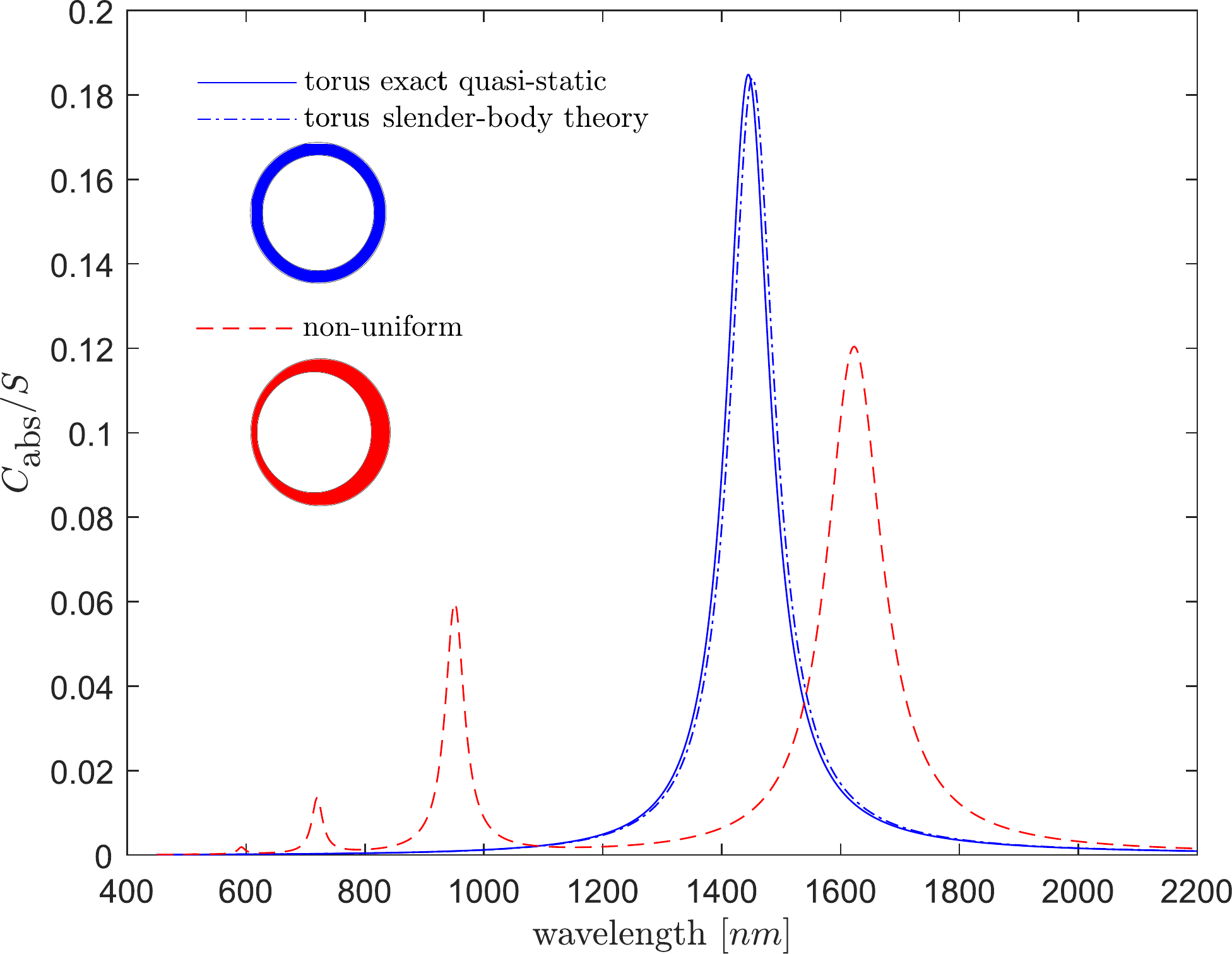}
	\caption{Absorption cross section \eqref{C def}, normalized by $S=\pi a^2$, for the torus-shaped and azimuthally non-uniform rings defined in the caption of Fig.~\ref{fig:single ring modes comparison}. The incident-field direction $\unit$ is in the plane of the ring, pointing in the maximum-thickness direction in the case of the azimuthally non-uniform ring. For the torus-shaped ring, the slender-body approximation \eqref{abs torus} is compared with an exact quasi-static solution \cite{Mary:07}. The slender-body approximation for the azimuthally nonuniform ring is evaluated numerically from \eqref{abs}. We assume the Drude model \eqref{drude} with $\omega_p= 1.196 \times 10^{16} \,\text{rad}/\text{s}$ and $\gamma = 8.05\times 10^{13}\,\text{rad}/\text{s}$.}
	\label{fig:absorption cross section single}
\end{figure}

\begin{figure}[]
	\centering\includegraphics[scale=0.6]{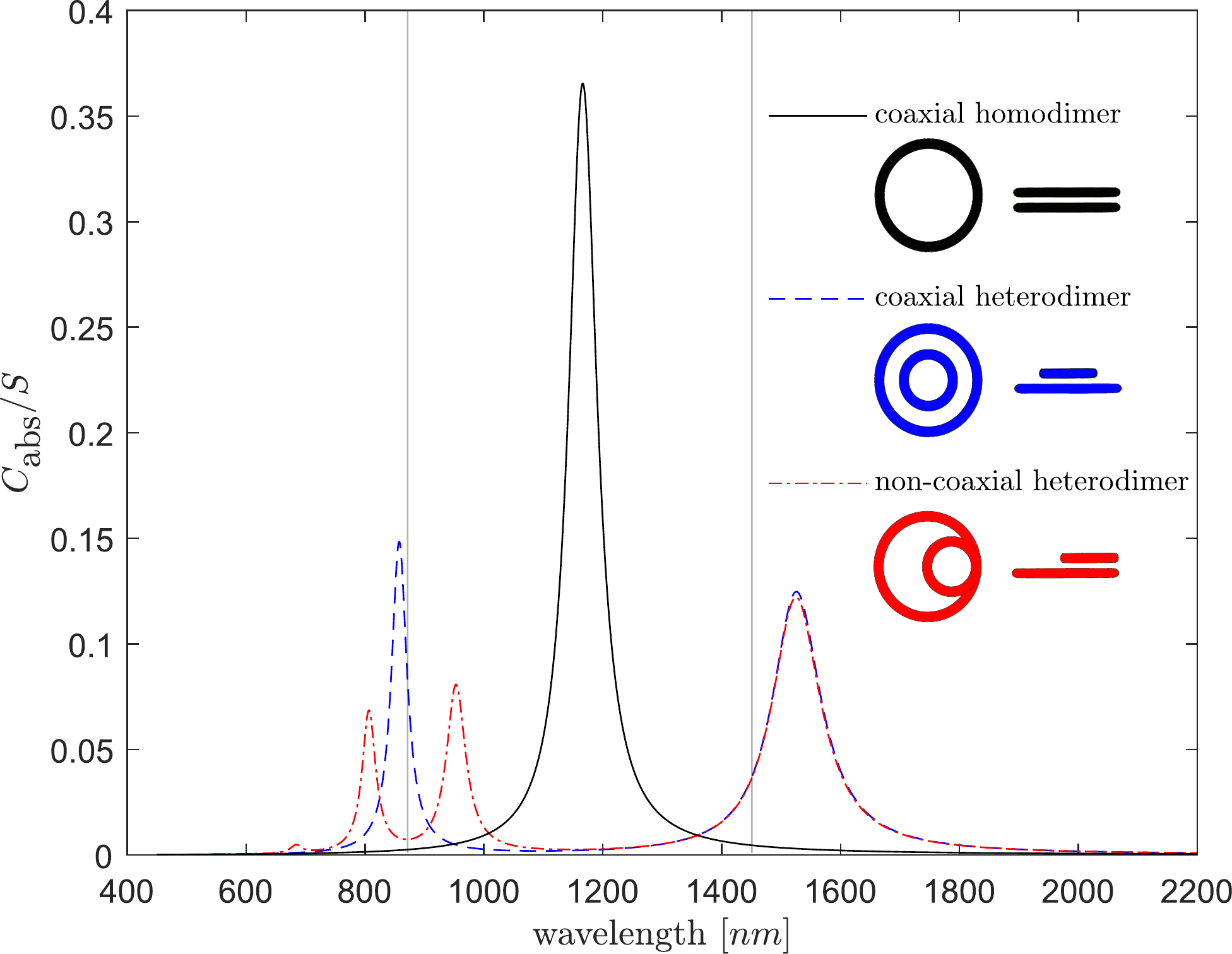}
	\caption{Absorption cross section \eqref{C def}, normalized by $S=\pi a_1^2$, for the  coaxial homodimer of Fig.~\ref{fig:co-axial dimer modes}; the coaxial heterodimer of Fig.~\ref{fig:co-axial dimer modes} and Fig.~\ref{fig:heterogeneous mode evol.} for $d=0$; and the bilayer heterodimer of Fig.~\ref{fig:heterogeneous mode evol.} for $d/a_1=0.5$. In all cases $a_1\ge a_2$ and the incident-field direction $\unit$ is in the plane of the rings, pointing in the direction of the horizontal displacement of the smaller ring in the non-coaxial case. The curves represent slender-body approximations evaluated using  \eqref{abs homogeneous co-axial dimer} for the coaxial homodimer, \eqref{abs co-axial dimer} for the coaxial heterodimer, and \eqref{abs} for the non-coaxial heterodimer. The vertical lines mark the  frequencies corresponding to the eigenvalues $\mathcal{E}\ub{1}$ of the larger and smaller torus-shaped rings in isolation. We assume the Drude model \eqref{drude} with $\omega_p = 1.196\times10^{16}\,\text{rad}/\text{s}$ and $\gamma = 8.05\times10^{13}\,\text{rad}/\text{s}$.} 	\label{fig:absorption cross section dimer}
\end{figure}
In Fig.~\ref{fig:absorption cross section single}, we show $C_{\text{abs}}$, normalized by the planar area $\pi a^2$, for a torus-shaped ring alongside that for the azimuthally non-uniform ring from Fig.~\ref{fig:single ring modes comparison}. The incident-field direction $\unit$ is parallel to the plane of the ring; in the case of the azimuthally non-uniform ring, $\unit$ is further specified as pointing towards the direction of maximum thickness of the ring. For the torus-shaped ring, the slender-body approximation \eqref{abs torus} is compared with the exact quasi-static computations in \cite{Mary:07} and good agreement is found despite the aspect ratio $\kappa=10$ being only moderately large. The torus-shaped ring exhibits a single resonance peak, which is associated with the excitation of the cosine-dipolar mode $(1,c)$ shown at the top-right corner of Fig.~\ref{fig:single ring modes comparison}. (With $\phi$ measured from $\unit$, the corresponding sine mode is not excited.)  In contrast, for the azimuthally non-uniform ring, we observe multiple resonance peaks. As explained in \S\ref{sssec:nonuniform}, in that case all modes generally include a dipolar component (first Fourier harmonic). Accordingly, the overlap integrals in \eqref{abs} do not vanish identically as they do in the torus case for $m>1$. In particular, the lowest-  and second-lowest-frequency peaks seen in Fig.~\ref{fig:absorption cross section single} are due to excitation of the `dipolar' mode $(1,c)$ and `quadrupolar' mode $(2,c)$ shown on the left of Fig.~\ref{fig:single ring modes comparison}.

In Fig.~\ref{fig:absorption cross section dimer}, we show $C_{\text{abs}}$, normalized by the planar area $\pi a_1^2$, for several bilayer dimer configurations formed of torus-shaped rings: a coaxial homodimer (same as in Fig.~\ref{fig:co-axial dimer modes}), a coaxial heterodimer (same as in Fig.~\ref{fig:co-axial dimer modes} and Fig.~\ref{fig:heterogeneous mode evol.} for $d=0$) and a bilayer non-coaxial heterodimer (same as in Fig.~\ref{fig:heterogeneous mode evol.} for $d/a_1=0.5$). The incident-field direction $\unit$ is taken parallel to the planes of the rings; in the non-coaxial-heterodimer case, $\unit$ is further specified as pointing along the direction of the horizontal displacement of the smaller ring. 

For the coaxial homodimer, we observe a single resonant peak, a prediction which is clearly inferred from the closed-form slender-body approximation \eqref{abs homogeneous co-axial dimer}. This is because only the in-phase cosine-dipolar mode $(1,c,+)$ is excited by the plane wave. (With $\phi$ measured from $\unit$, the corresponding sine mode is not excited.) While this is not demonstrated here, we know from the analytical expression \eqref{eigenvalue dimer same size} for $\mathcal{E}\ub{1,+}$, or from Fig.~\ref{fig:co-axial dimer modes}a, to expect this peak to blueshift with decreasing distance between the rings. This is opposite to the trend observed for the analogous low-frequency bonding-gap modes of sphere dimers \cite{Klimov:07,Pendry:13,Schnitzer:15plas, Schnitzer:18}. 

For the coaxial heterodimer, we observe an additional resonance peak at higher frequency. This is because the net dipole induced in the rings for the out-of-phase dipolar modes does not cancel out between the two rings as it does in the coaxial-homodimer case. This feature is easily inferred from the slender-body approximation \eqref{abs co-axial dimer}, which explicitly shows two resonances associated with the in-phase and out-of-phase dipolar modes. (With $\phi$ measured from $\unit$, it is again only the `cosine' dipolar modes that are excited.) As seen in Fig.~\ref{fig:co-axial dimer modes}b, for this configuration the out-of-phase dipolar mode is dominated by a dipolar distribution in the larger ring, whereas the in-phase dipolar mode is dominated by a dipolar distribution in the smaller ring. From Fig.~\ref{fig:co-axial dimer modes}a, we expect the in-phase and out-of-phase resonance to respectively blueshift and redshift as the vertical distance between the rings is reduced. 

Consider next the non-coaxial-heterodimer configuration. Similarly to the azimuthally non-uniform ring, we observe multiple resonance peaks owing to the absence of axial symmetry. In particular, we note that the higher-frequency resonance in the coaxial-heterodimer case, which is associated with the in-phase-dipolar mode $(1,c,+)$ of that configuration, is replaced by two distinct peaks. This can be understood from the study carried out in Fig.~\ref{fig:heterogeneous mode evol.}. Thus, this pair of resonances are associated with the excitation of the two modes continuated from the in-phase-dipolar $(1,c,+)$ and out-of-phase-quadrupolar $(2,c,-)$ modes of the corresponding coaxial-heterodimer configuration. While, owing to symmetry, in the coaxial case only dipolar modes are excited, in the non-coaxial case both modes include dipolar components and can therefore be excited. Specifically, the non-coaxial configuration is as shown in insets (b) and (g) of Fig.~\ref{fig:heterogeneous mode evol.}, in which case the eigenvalues are farthest apart as a function of the horizontal displacement, and both modes have a visible dipolar component in the smaller ring.

\section{Validity of quasi-static formulation}\label{sec:qsvalid}
The theory in this paper is predicated on the quasi-static approximation, which is widely used to describe the localised-surface-plasmon resonances of nanometallic structures \cite{Klimov:14}. In this approximation, radiation damping is neglected on the basis of the smallness of the particle relative to the free-space wavelength, and the near-field of the particle is assumed to be irrotational and forced  by the  field locally associated with the incident radiation. For sufficiently weak material loss, however, radiative corrections must in fact be appreciable even for particles which are much smaller than the free-space wavelength \cite{Aubry:10,Ruan:10}. In the concluding section \S\ref{sec:concluding}, we discuss opportunities to extend the present theory to incorporate radiative corrections for low-loss subwavelength particles, as well as to a full-wave theory for slender rings whose diameter is comparable to the free-space wavelength. Here, we derive a scaling condition for the quasi-static formulation to adequately describe the longitudinal resonances of a slender ring. Furthermore, we illustrate these constraints by  comparing with full-wave simulations.

\begin{figure}[]	\centering\includegraphics[scale=0.75]{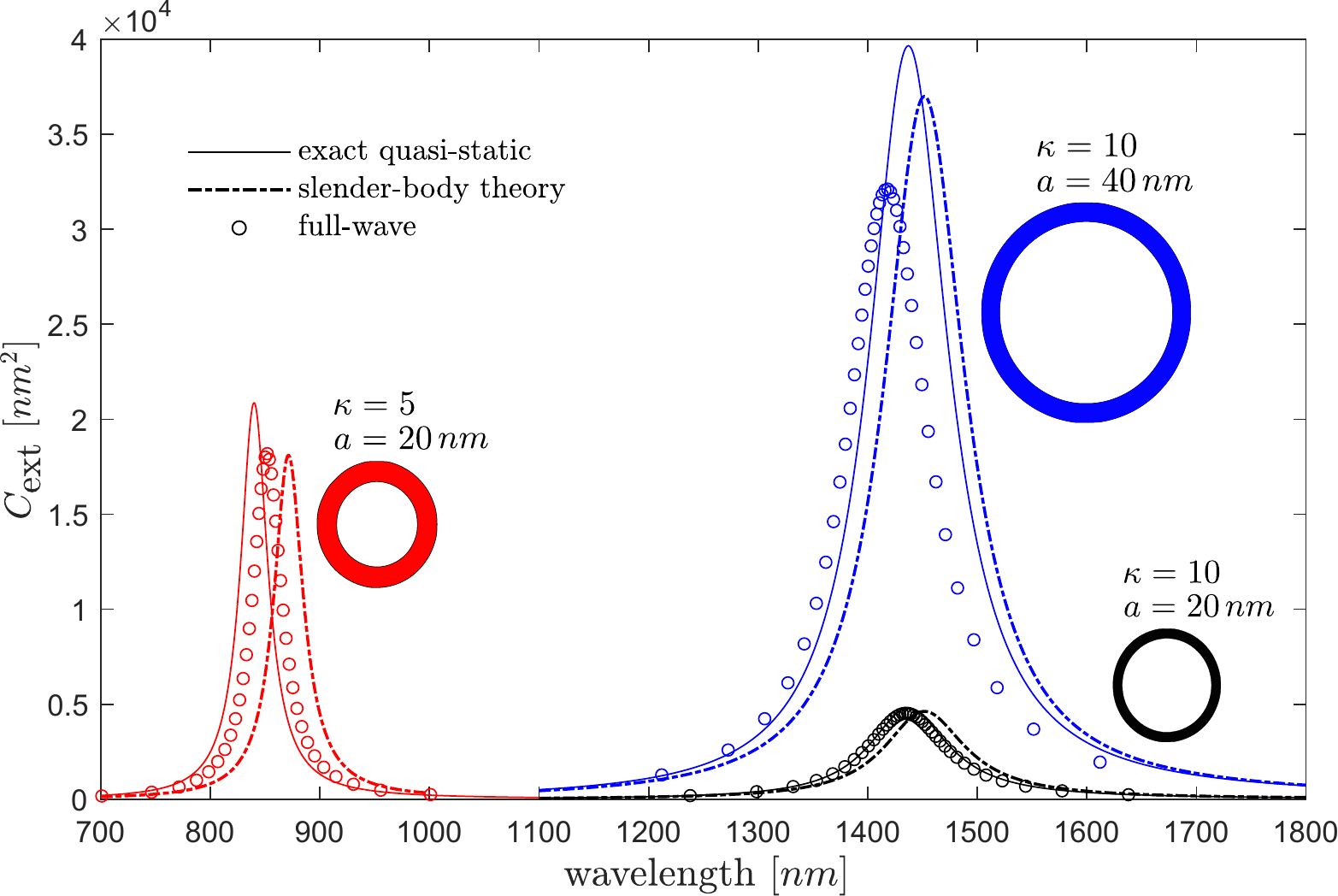}
	\caption{Extinction cross section for torus-shaped rings for the indicated values of the ring radius $a$ and aspect ratio $\kappa$. The incident field is parallel to the ring plane. Solid lines: solutions exact in the quasi-static approximation  \cite{Mary:07}. Dash-dotted lines: slender-body approximation \eqref{abs torus}. Circles: full-wave numerical computations using the MNPBEM toolbox \cite{MNPBEM}. We assume the Drude model \eqref{drude} with $\omega_p= 1.196 \times 10^{16} \,\text{rad}/\text{s}$ and $\gamma = 8.05\times 10^{13}\,\text{rad}/\text{s}$.} 	
	\label{fig:absorption cross full-wave}
\end{figure}
Consider first the condition for a slender ring to be subwavelength at its longitudinal resonances. At the level of scalings, and ignoring logarithmic factors, \eqref{drude} and \eqref{scaling} give the free-space wavelength $\lambda\simeq \kappa \lambda_p$, where $\lambda_p=2\pi c/\omega_p$. Thus, the condition $a\ll \lambda$ implies $a\ll \lambda_p\kappa$, or, for the typical value $\omega_p=1.196\times 10^{16}\,\text{rad}/\text{s}$,  $a\ll \kappa \times 150\text{nm}$. Under this condition, which is met even for quite large slender rings, the field is approximately irrotational in the vicinity of the ring, whereby the quasi-static modes remain relevant to the expansion of that near field. Nonetheless, the quasi-static scattering problem formulated in \S\ref{sec:QS}, and in particular its spectral solution \eqref{Field expansion}, is  valid under a more stringent condition. Thus, the scattering cross section, which scales like $k^4\alpha^2$ in the quasi-static approximation, must be negligible compared to the absorption cross-section. Using \eqref{C def}, this implies $\alpha\ll \lambda^3$. Estimating $\lambda$ at the longitudinal resonances as before, and using \eqref{abs torus} (with \eqref{scaling}), we find $a^3/\mathrm{Im}(\epsilon_r)\ll \kappa \lambda_p^3$. More explicitly, using the Drude model \eqref{drude} to estimate $\mathrm{Im}(\epsilon_r)$ (with \eqref{scaling}) we can write this condition as
\begin{equation}
\frac{a}{\lambda_p}\ll \left(\frac{\gamma}{\omega_p}\right)^{1/3}\times\kappa^{4/3}.
\end{equation}
For typical parameters (same $\omega_p$ as above and $\gamma=8.05\times 10^{13}\,\text{rad}/\text{s}$), we have $a\ll \kappa^{4/3}\times 30\text{nm}$.

In Fig.~\ref{fig:absorption cross full-wave}, we show a comparison between quasi-static and full-wave predictions for the extinction cross-section of a torus-shaped ring ($\kappa=10$, typical values of $\omega_p$ and $\gamma$ as above), with the incident field in the ring plane. The full-wave results are obtained by boundary-element simulation using the the MNPBEM toolbox \cite{MNPBEM}. As for the quasi-static results, both exact \cite{Mary:07} and slender-body approximations are shown (cf.~\eqref{abs torus}). 
This comparison demonstrates the improved agreement between full-wave and quasi-static predictions with decreasing $a$ and increasing $\kappa$. 

\section{Concluding remarks}\label{sec:concluding}
We have developed an approximate theory to describe the longitudinal localized-surface-plasmon resonances of slender metallic nanorings of virtually arbitrary shape, as well as more involved nanometallic structures formed of two or more such rings, the separation between the rings being comparable with their centerline radii. At the heart of the theory is an asymptotic reduction of the 3D plasmonic eigenvalue problem governing the material- and frequency-independent longitudinal modes of the structure to a reduced 1D formulation, in which the plasmonic eigenmodes are represented by azimuthal voltage and polarization-charge profiles attached to each ring. Once this reduced eigenvalue problem has been solved for a given geometry,  approximations for the near-field eigenfunctions in 3D can be extracted from their associated 1D eigenfunctions. When joined with the standard spectral theory of plasmonic resonance, this provides the means to obtain any quantity of interest in a physical scattering problem (that involves the same geometry), such as scattering cross sections. Overall, the present theoretical framework allows to rapidly gauge the plasmonic properties of unprecedentedly complex 3D structures, which may foster the study of new plasmonic structures and phenomena.

Our approximation approach heavily draws on a transfer of knowledge from the field of fluid dynamics, where approximations of the sort derived here are known as slender-body theory. We stress that the slender-body theory developed herein is of the more accurate `nonlocal' type, which in the present context means that we account for electrostatic interactions between any two azimuthal segments of any of the rings forming the structure. The inclusion of these interactions can be shown to ensure an `algebraic' rather than `logarithmic' asymptotic accuracy of the scheme in the large-aspect-ratio limit pertinent to slender rings. This claimed accuracy is demonstrated by the good agreement in Figs.~\ref{fig:eig_comparison} and \ref{fig:absorption cross section single} with known solutions for single torus-shaped rings. 

To demonstrate the versatility of the theory, we have applied it to the calculation of the longitudinal modes and absorption cross sections of a number of   configurations. Still, it is clear that we only considered a small sample of the wide range of geometries that can be studied. For several families of geometries, namely azimuthally invariant rings (not necessarily torus-shaped) and coaxial dimers and chains thereof, the reduced formulation was solved in closed form, thus generating several apparently new analytical approximations. We also considered a range of other geometries, including azimuthally nonuniform rings and non-coaxial multiple-ring configurations. In the latter cases, we solved the reduced formulation using straightforward semi-analytical schemes, in which the 1D eigenfunctions are represented by their Fourier coefficients. 

We have focused on the description of the longitudinal (low-frequency) modes of slender-ring structures, as defined in the introduction and in \S\ref{sec:reduced evp}, as well as the excitation of these modes by incident plane waves. Slender rings also possess non-longitudinal, or transverse modes, characterised by polarization-charge distributions which, in contrast to the longitudinal case, exhibit significant variation in the cross-sectional planes. As we discussed in \cite{Ruiz-Schnitzer:19}, and others have observed in theory and experiments \cite{Link:99bis,Murphy:2005}, the contribution of non-longitudinal modes to the resonant response of slender bodies is generally negligible, except in specialised excitation scenarios involving near-field sources located very close to the structure. To treat those cases, our slender-body theory could be extended to describe the transverse modes of slender-ring structures. A leading-order analysis, as done in \cite{Ruiz-Schnitzer:19} for axisymmetric bodies, shows that those modes essentially correspond to the in-plane modes of an infinite cylinder, hence with the permittivity eigenvalues satisfying the asymptotic behaviour $\mathcal{E}\rightarrow-1$ as $\kappa\to\infty$. (Alluding to Drude's model \eqref{drude}, we see that transverse modes are  high frequency relative to longitudinal modes.) To consider corrections to this leading behaviour, one would need to match the dipolar (or quadruople, etc.) cross-sectional potential distributions with an approximation of the potential on the ring scale formed by superposition of the corresponding potential singularities along the centreline of the body.

Another desirable extension to this work would be to consider the effects of retardation and radiation damping. In particular, there exists a generalization of the spectral approach employed herein based on permittivity eigenvalues to the full-wave regime \cite{Bergman:80,Agranovich:99,Farhi:15,Chen:19}. In that generalized formulation, the permittivity eigenvalues become complex valued and frequency dependent. 
To describe the regime of low-loss and moderately subwavelength slender-ring structures, where deviations from the quasi-static can be substantial (as demonstrated in \S\ref{sec:qsvalid}), it would be necessary to carry out a low-frequency analysis of the generalized eigenvalue problem up to the order of the permittivity eigenvalue that is comparable to the imaginary part of the physical permittivity. (This regime could also be described by applying suitable corrections to the quasi-static predictions following \cite{Wokaun:82,Draine:94,Carminati:06,Aubry:10}.) Another avenue would be to develop a full-wave slender-body theory in the case where the ring diameter is comparable to the free-space wavelength; the cross-sectional scale would still be subwavelength, while the ring-scale approximation could be formed by superposition of fundamental solutions of Maxwell's equations along the centerline of the ring. Lastly, we note that it is possible but not straightforward to generalize the  permittivity-eigenvalue formulation to multi-constituent photonic structures \cite{Bergman:20}, e.g., hybrid metal-dielectric structures. For such ring structures, a reformulation of the theory based on quasi-normal modes and complex frequency eigenvalues may be more convenient \cite{Sauvan:13}. 

\textbf{Acknowledgement}. The authors acknowledge funding from EPSRC UK (New Investigator Award EP/R041458/1).

\appendix
\section{Matching the ring and cross-sectional scales}\label{app:matching}
In this appendix, we shall derive the small $r$ behavior of the ring-scale potential \eqref{outer}. A preliminary step is to write that solution in the form
\begin{equation}\label{outer potential proof}
		\varphi =  \frac{q(\phi)}{4\pi\epsilon_0}\int_{0}^{2\pi}\,a d\phi'\, \frac{1}{|\mathbf{x}(r,\theta,\phi)-\mathbf{y}(\phi')|} +\frac{1}{4\pi\epsilon_0} \int_{0}^{2\pi}\,ad\phi'\,\frac{q(\phi')-q(\phi)}{|\mathbf{x}(r,\theta,\phi)-\mathbf{y}(\phi')|},
	\end{equation}
where the geometry gives
	\begin{gather}\label{x-X}
		|\mathbf{x}(r,\theta,\phi)-\mathbf{y}(\phi')| = \sqrt{r^2 +  4(a^2+ar\cos\theta)\sin^2\frac{\phi-\phi'}{2}}.  
	\end{gather}
	Using \eqref{x-X}, it is readily seen that the second integral in \eqref{outer potential proof} is regular as $r\rightarrow0$: 
	\begin{equation}\label{integral eval. regular}
	\int_{0}^{2\pi}\,ad\phi'\,\frac{q(\phi')-q(\phi)}{|\mathbf{x}(r,\theta,\phi)-\mathbf{y}(\phi')|}= \int_{0}^{2\pi}\,d\phi'\, \frac{q(\phi')-q(\phi)}{2\sin\frac{\left|\phi-\phi'\right|}{2}} + o(1) \quad\text{as}\quad r\rightarrow0.
	\end{equation}

To treat the first integral in \eqref{outer potential proof}, which is singular as $r\to0$, we use the method of splitting the range of integration \cite{Hinch:91}. Thus, we write
\begin{equation}\label{splitted}
\int_{0}^{2\pi}\,a d\phi'\, \frac{1}{|\mathbf{x}(r,\theta,\phi)-\mathbf{y}(\phi')|} =\left\{\int_0^\delta\,dt+\int_{\delta}^{\pi}\,dt\right\}\frac{2}{\sqrt{\bar{r}^2+4(1+\bar{r}\cos\theta)\sin^2\frac{t}{2}}},
\end{equation}
where we introduce the normalized radial coordinate $\bar{r}=r/a$ and  an auxiliary parameter in the range $\bar{r}\ll\delta\ll1$. Denote the first integral on the right-hand side of \eqref{splitted} by $I_1$. Making the change of variables $\tau=t/\bar{r}$, we have
\begin{equation}
I_1=2\int_0^{\delta/\bar{r}} \frac{d\tau}{\sqrt{1+4(1+\bar{r}\cos\theta)\bar{r}^{-2}\sin^2(\bar{r}\tau/2)}}.
\end{equation}
Using $\bar{r}\tau<\delta\ll1$, the integrand can be expanded to show that
\begin{equation}\label{I1 result}
I_1  = 2\int_0^{\delta/\bar{r}}\frac{d\tau}{(1+\tau^2)^{1/2}}+o(1) = 2\ln \frac{2\delta}{\bar{r}} +o(1)
  \quad \text{as} \quad \bar{r}\to0,
\end{equation}
where for the last step we used $\delta/\bar{r}\gg1$. Similarly, denote the second integral on the right-hand side of \eqref{splitted} by $I_2$. Using $t>\delta\gg\bar{r}$, the integrand of that integral can be expanded to show that
\begin{equation}\label{I2 result}
I_2 = \int_\delta^\pi\frac{dt}{\sin \frac{t}{2}}+o(1) = 2\ln \frac{4}{\delta}+o(1) \quad \text{as} \quad \bar{r}\to0. 
\end{equation}
Combining \eqref{integral eval. regular}, \eqref{I1 result} and \eqref{I2 result}, we find the requisite behavior
\begin{equation}\label{outer to inter}
		\varphi(\bx) = \frac{q(\phi)}{2\pi\epsilon_0}\ln\frac{8a}{r}+\frac{1}{4\pi\epsilon_0}\int_0^{2\pi}d\phi'\,\frac{q(\phi')-q(\phi)}{2\sin\frac{|\phi-\phi'|}{2}} + o(1)\quad\text{as}\quad r\to 0.
\end{equation}

\section{Proof of identity \eqref{int identity}}
\label{app:identity}
In this appendix we prove identity \eqref{int identity}, which constitutes a diagonalization   of the integral operator appearing in the capacitance relation \eqref{cap}. To this end, it is convenient to denote that integral operator as
	\begin{equation}\label{I def}
		\mathcal{N}[q(\phi)] = \int_{0}^{2\pi}d\phi'\, \frac{q(\phi')-q(\phi)}{2\sin\frac{\left|\phi-\phi'\right|}{2}}.
\end{equation}

We shall show that, for any non-negative integer $n$, there exists a constant $\lambda_n$ such that
\refstepcounter{equation}
$$
\label{eigenvalue int operator appendix}
	\mathcal{N}[\cos n\phi] = \lambda_n\cos n\phi, \quad     \mathcal{N}[\sin n\phi ] = \lambda_n\sin n\phi.
\eqno{(\mathrm{\theequation}{\mathrm{a},\!\mathrm{b}})}
$$
Our method of proof will also furnish the constants, which constitute the eigenvalues of the operator $\mathcal{N}$. In what follows, we consider only the cosine eigenfunctions, the corresponding result for the sine eigenfunctions readily following from a straightforward change of variables.  

Direct calculation shows that $\mathcal{N}[1] = 0$ and  $\mathcal{N}[\cos \phi] = -4\cos \phi$, thence that (\ref{eigenvalue int operator appendix}a) holds for $n=0$ and $n=1$ with $\lambda_0=0$ and $\lambda_1=-4$. 
	Now, let $m$ be any integer $>1$ and assume that (\ref{eigenvalue int operator appendix}a) also holds for $n=1,\ldots,m-1$, with 
\begin{equation}\label{difference assumption}
\lambda_n=-\frac{4}{2n-1}+\lambda_{n-1}.
\end{equation}
This assumed difference relation is consistent with the  result for $n=1$ and implies that
\begin{equation}\label{lambda n step}
\lambda_n= -4\sum_{k=1}^n\frac{1}{2k-1} \quad \text{for} \quad n=1,\ldots,m-1.
\end{equation}
Next, we use the definition \eqref{I def} and (\ref{eigenvalue int operator appendix}a) for $n=m-2$ to write 
\begin{eqnarray}\label{int operator proof 1}
	\mathcal{N}[\cos m\phi]     &=& 2\mathcal{N}[\cos\left((m-1)\phi\right)\cos \phi] - \lambda_{m-2}\cos\left((m-2)\phi\right).
\end{eqnarray}
Using \eqref{I def} (\ref{eigenvalue int operator appendix}a) for $n=m-1$, we have
	\begin{multline}\label{int operator proof 2}
		\mathcal{N}[\cos\left((m-1)\phi\right)\cos\phi]  =\int_{0}^{2\pi}d\phi'\,\frac{\cos\left((m-1)\phi'\right)\cos\phi' - \cos\left((m-1)\phi\right)\cos\phi}{2\sin\frac{|\phi' - \phi|}{2}}\\ 
		= \int_{0}^{2\pi}d\phi'\,\frac{\cos\left((m-1)\phi'\right)\left(\cos \phi' -\cos\phi\right)}{2\sin\frac{|\phi' - \phi|}{2}} + \cos\phi\,\mathcal{N}[\cos\left((m-1)\phi\right)]\\
		= -\int_{0}^{2\pi}d\phi'\,\frac{\cos\left((m-1)\phi'\right)\sin{\frac{\phi'-\phi}{2}}\sin{\frac{\phi'+\phi}{2}}}{\sin\frac{|\phi' - \phi|}{2}} + \lambda_{m-1}\cos\phi \cos\left((m-1)\phi\right)\\
		= 2\frac{\cos((m-2)\phi)}{2m-3} - 2\frac{\cos m\phi}{2m-1} + \frac{1}{2}\lambda_{m-1}\left(\cos m\phi + \cos((m-2)\phi)\right).
	\end{multline} 
So that combining \eqref{int operator proof 1} and \eqref{int operator proof 2} gives
\begin{gather}
	\mathcal{N}[\cos m\phi] =  \left(\lambda_{m-1}-\frac{4}{2m-1} \right)\cos m\phi + \left( \lambda_{m-1} - \lambda_{m-2}+\frac{4}{2m-3}\right)\cos((m-2)\phi),
\end{gather}
where the second term on the right-hand side vanishes owing to \eqref{difference assumption}, while the first term on the right-hand side shows that \eqref{difference assumption}, and thence \eqref{lambda n step}, holds also for $n=m$. 
It follows by induction that \eqref{eigenvalue int operator appendix} holds for any non-negative integer $n$, with $\lambda_0=0$ and $\lambda_n$ given by \eqref{lambda n step} for $n>0$. 

\section{Semi-analytical scheme}
\label{app:scheme}
\subsection{Single ring} \label{app:scheme single}
For a single ring of non-uniform thickness, the reduced eigenvalue problem of \S\ref{sec:reduced evp} is approximated by the generalized matrix-eigenvalue problem \eqref{numerical scheme}. In that problem, $\mathsf{M} = \{M_{n,k}\}$ and $\mathsf{U} = \{U_{n,k}\}$ are $2K\times 2K$ matrices with components 
\begin{equation}\label{scheme matrix 1}
	M_{n,k} = \frac{nk}{\pi} \times 
	\begin{dcases}
		 \int_0^{2\pi}d\phi\,\bar{A}(\phi)\sin n\phi \sin k\phi, \quad   n \leq K,\, k\leq K\\	
		-\int_0^{2\pi}d\phi\,\bar{A}(\phi)\sin n\phi\cos k\phi, \quad n \leq K,\, k> K\\		
		-\int_0^{2\pi}d\phi\,\bar{A}(\phi)\cos n\phi \sin k\phi, \quad  n > K,\, k\leq K\\		 
		\int_0^{2\pi}d\phi\,\bar{A}(\phi)\cos n\phi\cos k\phi, \quad n > K,\, k > K
	\end{dcases}
\end{equation}
and
\begin{equation}\label{scheme matrix 2}
	U_{n,k} = -\frac{1}{\pi}\delta_{n,k}\sum_{k=1}^n\frac{1}{2k-1} +\frac{1}{2\pi^2}\times
	\begin{dcases}
		\int_0^{2\pi}d\phi\,\cos n\phi\cos k\phi\ln{\frac{8\kappa}{f(\phi)}}, \quad n \leq K,\, k\leq K\\		
		\int_0^{2\pi}d\phi\,\cos n\phi\sin k\phi\ln{\frac{8\kappa}{f(\phi)}}, \quad n \leq K,\, k> K\\		
		\int_0^{2\pi}d\phi\,\sin n\phi\cos k\phi\ln{\frac{8\kappa}{f(\phi)}}, \quad n > K,\, k\leq K\\		
		\int_0^{2\pi}d\phi\,\sin n\phi\sin k\phi\ln{\frac{8\kappa}{f(\phi)}}, \quad n > K,\, k > K.
	\end{dcases}
\end{equation}
Furthermore, the coefficient $\alpha_0$ can be calculated \textit{a posteriori} as
\begin{equation}
	\alpha_0 = \frac{1}{4\pi^2}\sum_{k=1}^{K}\int_0^{2\pi}d\phi\,\left(\tilde{\alpha}_k\cos k\phi + \tilde{\beta}_k\sin k\phi\right)\ln{\frac{8\kappa}{f(\phi)}}.
\end{equation}

\subsection{Ring dimer}\label{app:scheme dimer}
For an arbitrary ring dimer, the generalized eigenvalue problem of \S\ref{ssec:reduced2} is approximated by the generalized matrix-eigenvalue problem 
\eqref{numerical scheme app dimer}. In that problem, the $2K\times 2K$ matrices $\mathsf{M}_n$ and $\mathsf{U}_n$ are like above with $n=1,2$ indicated the ring number, whereas $\mathsf{V}=\{V_{n,k}\}$ is a $2K\times 2K$ coupling matrix with components
\begin{equation}
	V_{n,k} = \frac{\sqrt{a_1a_2}}{4\pi^2}
	\begin{dcases}
		\int_{0}^{2 \pi}\int_{0}^{2 \pi}d\phi_1 d\phi_2\,\frac{\cos n\phi_1 \cos k\phi_2}{|\mathbf{y}_1(\phi_1)-\mathbf{y}_2(\phi_2)|}, \quad n \leq K,\, k\leq K\\	
	\int_{0}^{2 \pi}\int_{0}^{2 \pi}d\phi_1 d\phi_2\,\frac{\cos n\phi_1 \sin k\phi_2}{|\mathbf{y}_1(\phi_1)-\mathbf{y}_2(\phi_2)|}, \quad n \leq K,\, k> K\\		
	\int_{0}^{2 \pi}\int_{0}^{2 \pi}d\phi_1 d\phi_2\,\frac{\sin n\phi_1 \cos k\phi_2}{|\mathbf{y}_1(\phi_1)-\mathbf{y}_2(\phi_2)|}, \quad n > K,\, k\leq K\\		 
	\int_{0}^{2 \pi}\int_{0}^{2 \pi}d\phi_1 d\phi_2\,\frac{\sin n\phi_1 \sin k\phi_2}{|\mathbf{y}_1(\phi_1)-\mathbf{y}_2(\phi_2)|}, \quad n > K,\, k > K.\\
	\end{dcases}
\end{equation} 

\section{Far-field expansion}\label{appendix far-field}
In this appendix we derive the far field behavior of the scattered field starting from the spectral solution \eqref{Field expansion}, with the eigenvalues and eigenmodes approximated by their slender-body approximations. As in \S\ref{sec:PW}, we only include longitudinal modes in this calculation. For generality, we consider a system of $N$ rings. 

Consider first the far-field behavior of the eigenpotentials $\varphi\ub{I}(\bx)$, whose slender-body approximations are  (cf.~\eqref{outer2}) 
\begin{equation}\label{outerN}
	\varphi\ub{I}(\bx) = \sum_{n=1}^N\frac{a_n}{4\pi\epsilon_0}\int_0^{2\pi} d\phi'\,\frac{q_n\ub{I}(\phi')}{|\bx-\mathbf{y}_n(\phi')|}.
\end{equation}
Using the asymptotic behavior
\begin{equation}\label{asymp 1/x}
	\frac{1}{|\mathbf{x} - \mathbf{y}_n|} \sim  \frac{1}{|\mathbf{x}|} + \frac{\mathbf{x}\boldsymbol{\cdot}\mathbf{y}_n}{|\mathbf{x}|^3} 
\quad\text{as}\quad|\mathbf{x}|\to\infty, 
\end{equation}
for $n=1,2,\ldots,N$, and the zero-charge constraints $\int_0^{2\pi}d\phi\,q_n\ub{\mathcal{I}} = 0$, \eqref{outerN} gives 
\begin{equation}
	\varphi\ub{\mathcal{I}}(\mathbf{x}) \sim  \frac{\mathbf{x}}{4\pi\epsilon_0|\mathbf{x}|^3}\boldsymbol{\cdot}\sum_{n=1}^Na_n\int_0^{2\pi}d\phi\,\mathbf{y}_n\,q_n\ub{\mathcal{I}} \quad\text{as}\quad|\mathbf{x}|\to\infty.
\end{equation}

We next note that, for the sake of evaluating the overlap and normalization integrals in \eqref{Field expansion}, we may approximate the eigenfield inside the $n$th ring as
\begin{equation}
\boldsymbol{\nabla}\varphi_n\ub{\mathcal{I}}= \frac{1}{a_n^2}\frac{dv_n\ub{I}}{d\phi}\pd{\mathbf{y}_n}{\phi}.
\end{equation}
Thus, using integration by parts together with the effective Gauss law \eqref{Gauss coupled}, we find 
\begin{equation}
	\frac{\int dV\, \mathbf{E}_{\infty}
		\boldsymbol{\cdot} \boldsymbol{\nabla}\varphi\ub{I}}{\int dV\, \boldsymbol{\nabla}\varphi\ub{I} \boldsymbol{\cdot} \boldsymbol{\nabla}\varphi\ub{I}} = \frac{\sum_{n=1}^N a_n\int_0^{2\pi}d\phi\,
		\mathbf{E}_{\infty}\boldsymbol{\cdot}\mathbf{y}_n\, q_n\ub{I}}{\sum_{n=1}^N a_n\int_0^{2\pi}d\phi\,
		v_n\ub{I} q_n\ub{I}}.
\end{equation}
We emphasize that this result is valid even for azimuthaly non-uniform rings and arbitrary cross-sectional shapes. 

With the above slender-body approximations, the spectral solution \eqref{Field expansion} gives the far-field behavior \eqref{far field generic}, with the polarization tensor $\boldsymbol{\alpha}$ given by \eqref{abs}.

\bibliographystyle{apsrev4-2}
\bibliography{ref}

\end{document}